\documentclass[showpacs,aps,prb,twocolumn]{revtex4}
\usepackage{amsmath,amsthm,amsfonts}  
\usepackage{epsf}
\usepackage{amssymb}  
\usepackage{graphicx}
\def\Xint#1{\mathchoice
   {\XXint\displaystyle\textstyle{#1}}%
   {\XXint\textstyle\scriptstyle{#1}}%
   {\XXint\scriptstyle\scriptscriptstyle{#1}}%
   {\XXint\scriptscriptstyle\scriptscriptstyle{#1}}%
   \!\int}
\def\XXint#1#2#3{{\setbox0=\hbox{$#1{#2#3}{\int}$}
     \vcenter{\hbox{$#2#3$}}\kern-.5\wd0}}

\def\dashint{\Xint-}

\begin{document}

\title{Adiabatic transfer of electrons in coupled quantum dots}

\author{T. Brandes $^1$ and T. Vorrath $^2$}
\affiliation{ $^1$ Department of Physics, University of Manchester Institute of
Science and Technology (UMIST), P.O. Box 88, Manchester M60 1QD, United 
Kingdom\\
$^2$Univ. of Hamburg, 1. Inst. Theor. Physik, Jungiusstr. 9, D-20355 Hamburg, Germany}

\date{\today{ }}

\begin{abstract}
We investigate the influence of dissipation on one-- and two--qubit rotations in coupled semiconductor
quantum dots, using a (pseudo) spin--boson model with adiabatically varying parameters.
For weak dissipation, we solve a master equation, compare with direct perturbation theory,
and derive an expression for the `fidelity loss' during a simple operation that adiabatically 
moves an electron between two coupled dots. 
We discuss the possibility of visualizing 
coherent quantum oscillations in electron `pump' currents, combining quantum 
adiabaticity and Coulomb blockade.
In two--qubit spin--swap operations where the role of intermediate charge states 
has been discussed recently, we apply our formalism to calculate the 
fidelity loss due to charge tunneling between two dots. 
\end{abstract}
\maketitle

\section{Introduction}
The control of quantum superpositions in coupled electronic systems
has been suggested as a possible way to realize quantum logic gates in semiconductor structures. 
An example are coupled quantum dots \cite{Vaartetal95,Blietal96,Fujetal98,Blietal98b,Taretal99}, where 
Coulomb interactions between electrons can be exploited \cite{LD98,BLD99,BL00} to 
define very small effective Hilbert spaces  such as that of two tunnel-splitted  
ground states, separated by a large energy gap from the remaining many-particle states\cite{SN96,BK99,HD00}.

If parameters of the Hamiltonian (like the tunnel-coupling between dots)
are slowly changed as a function of time, adiabatic control of the state vector \cite{Bonetal98},
swap operations \cite{SLM01}, and the controlled transfer from an initial to a final
state \cite{BRB01} become possible. 
In addition, by coupling such a  system 
to external electronic reservoirs, one can pump  
electrons through the system in a controlled, adiabatic  manner which in principle can 
serve as a `read out' of the state vector in form of an electric current.
In the simplest case,
this can be achieved through a simultaneous variation of two parameters as a function of time.

Both adiabatic control of rotations or swaps, and coherent pumping of electrons evidentally are 
very sensitive to decoherence.
The coupling to dissipative degrees of freedom such as phonons disturbs the coherent
time-evolution and therefore leads to a loss of control over the desired superposition.
Decoherence and dissipation in quantum XOR gate operations have been discussed recently by 
Thorwart and  H\"anggi \cite{TH02} in a powerful numerical scheme. They found that properties like
gate fidelities are very sensitive to the dissipative bath coupling constant, but
only weakly depend on temperature. 

In this article, we quantitatively investigate the role of dissipation for 
one- and two-qubit adiabatic `rotations' of an electron between two
coupled quantum dots. For the electron charge one-qubit, we discuss a `quantum electron pump' 
for dots which are coupled to external leads (electron reservoirs).
In our scheme, decoherence is mainly due to absorption of bosons in excitations
of the instantaneous ground state which leads to an exponential temperature
dependence. 
For a sinusoidal pulse, we use an exact solution and perturbation theory in
the phonon coupling to predict how quantum mechanical oscillations between the dots can be made 
visible in a `read out' electronic current,  similar to the experiment
by Nakamura {\em et al.} in a superconducting Cooper pair box \cite{NPT99} and 
the recent electron spin resonance scheme for dots by Engel and Loss \cite{EL01}.

First, we compare results for both weak and strong 
coupling to the bosons and derive an analytic expression for the influence of 
the bath on the fidelity of the operation. 
Second,  we investigate two-qubit swap operations  in spin-based  two-electron quantum dots.
These {\em spin} qubits sensitively depend on {\em charge} decoherence due to  intermediate states 
where charge has tunneled between the dots. Piezoelectric phonons coupled to the electron charge 
incoherently mix states in the singlet sector and lead to a loss of fidelity of the swap operation.

Adiabatic transfer and pumping of 
charges through small metallic islands or semiconductor quantum dots 
\cite{Geretal90,Kouetal91,Potetal92,Grabert} has already been demonstrated experimentally.
Furthermore, in the strong Coulomb blockade regime of coupled quantum dots, 
the non-adiabatic coupling to AC fields in photo-assisted transport \cite{SW96,BBS97,SN96,SWL00}
has been established in experiments \cite{Oosetal98,Blietal98a,Holetal00}.
In the opposite regime of weak Coulomb correlations, 
experiments in open dots \cite{Swietal99} have demonstrated the feasability of 
an `adiabatic quantum electron pump'. 
These systems can be described as  non-interacting  mesoscopic scatterers
\cite{Bro98,PB01,CB02,MB01,MCM02}. 

A {\em combination} of strong Coulomb blockade and the adiabatic control of the wave function
in a triple quantum dot has been suggested recently \cite{RB01}.
Furthermore, in superconducting Josephson junction qubits \cite{MSS01},
adiabatic quantum computation with Cooper pairs
\cite{Ave98} and adiabatic controlled-NOT gates for quantum computation have been 
proposed by Averin \cite{Ave99}. 

The decoherence properties of adiabatic one- and two-qubit operations 
are closely related to the dissipative Landau-Zener problem (see, e.g., the review
\cite{GH98} for further references). 
In time-dependent operations, decoherence rates in general 
become time-dependent themselves and therewith usually involve the whole 
(or a large part) of the spectrum of frequencies in  the 
effective spectral density $\rho(\omega)$ of the bosonic bath.
As we show below, adiabatic pulses 
can be chosen such that qubit rotations run at a {\em constant} energy gap $\Delta$ to the excited state
whereby dissipation is due to $\rho(\omega=\Delta/\hbar)$, i.e., one boson frequency only. 
This can be used to extract $\rho(\omega)$ from the pump current by using a series
of pulses with different $\Delta$. 

A suppression of $\rho(\omega)$ 
at certain frequencies $\omega=\omega_0$ has been predicted in free-standing
`phonon cavities' due to symmetry and geometrical confinement \cite{DBK02}.
Here, we demonstrate that a `Rabi' rotation
pulse tuned to constant energy difference $\Delta(t)=\hbar\omega_0$ effectively
`switches off' the decoherence in such systems, at least within second order in the coupling constant.
Consequently, this defines a 
one-dimensional `decoherence-free manifold' (curve) on the adiabatic groundstate energy surface
of the system, which might be of interest for adiabatic quantum computation schemes
suggested recently \cite{CFP01}.

The paper is organized as follows: in section \ref{section_one_qubit}, we 
discuss one-qubit rotations, introduce the time-dependent spin-boson model, and
describe the adiabatic transfer of an electron between quantum dots with and without
dissipation. In section \ref{sectionfidelity}, we derive a perturbative, analytical
expression for the one-qubit fidelity, compare with the strong-coupling case, and suggest
a scheme to extract quantum oscillations as a `read out' electron pump current. 
In section \ref{section_two_qubit}, we turn to spin-qubit swaps and charge decoherence for the 
gate discussed recently by Schliemann, Loss, and MacDonald.
Finally, section \ref{conclusion} is a short conclusion.

\section{One-Qubit Rotations}\label{section_one_qubit}
Adiabatic transfer in a two-level system consists in the rotation of a chosen  
initial state $|{\rm in}\rangle$ into a final state $|{\rm out}\rangle$ by an adiabatic variation of 
system parameters. The simplest example is a spin $\frac{1}{2}$ in a slowly rotating magnetic
field. 

\begin{figure}[t]
\includegraphics[width=0.2\textwidth]{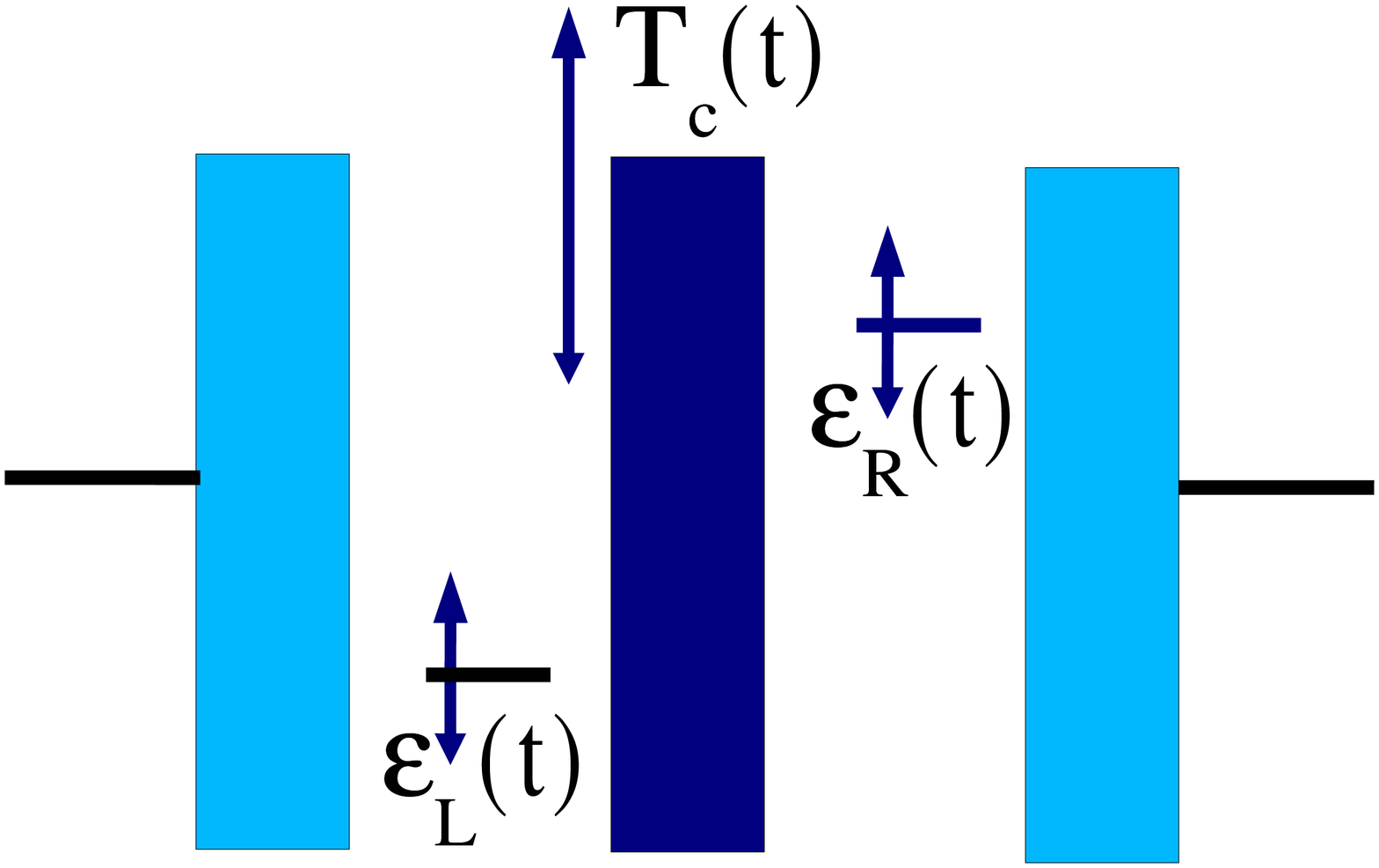}
\includegraphics[width=0.2\textwidth]{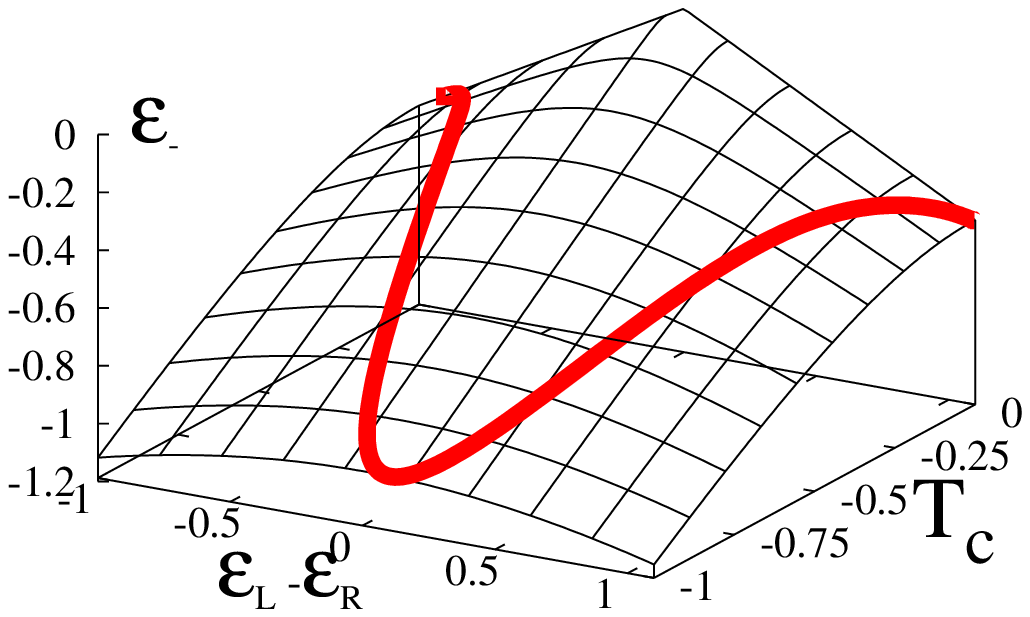}
\caption[]{\label{2surface}Left: Double dot  with time-dependent energy level difference
$\varepsilon(t)=\varepsilon_L(t)-\varepsilon_R(t)$ and
tunnel matrix element $T_c(t)$, connected to electron reservoirs.
Right:
Surface of the lower energy eigenvalue  $\varepsilon_-$ of the two-level Hamiltonian 
$H_0^{(1)}(t)$, Eq. (\ref{H0define}).
To adiabatically transfer an electron from the left to the right dot, $\varepsilon$ and 
$T_c$ are varied as a function of time as in Eq. (\ref{etpulse}),
corresponding to the curve on the $\varepsilon_-$ surface. }
\end{figure}

Using the analogy with adiabatic steering in atomic three-level systems, 
the possibility of adiabatic pumping of electrons through {\em triple} quantum dots 
has been suggested recently \cite{RB01}, using a suitable
`design' of instantaneous energies through two time-dependent tunnel couplings. 
Here, we consider the transfer of an electron 
through a {\em double} quantum dot via quasi-stationary adiabatic eigenstates.
This `charge qubit' \cite{BL00}, Fig. (\ref{2surface}) left,  is defined by two electron states
$|L\rangle$ and $|R\rangle$ with a time-dependent energy difference $\varepsilon(t)=
\varepsilon_L(t)-\varepsilon_R(t)$
and coupled by a tunnel matrix element $T_c(t)$, as 
described by the time-dependent Hamiltonian 
\begin{eqnarray}\label{H0define}
  H_0^{(1)}(t) &=& \frac{\varepsilon(t)}{2}\sigma_z + T_c(t) \sigma_x, 
\end{eqnarray}
with $\sigma_z:=|L\rangle\langle L| - |R\rangle\langle R|$ and 
$\sigma_x:=|L\rangle\langle R| + |R\rangle\langle L|$. 
Experimental control of (constant) $\varepsilon$ and $T_c$ has been 
demonstrated in double quantum dots \cite{Vaartetal95,Blietal98b,Fujetal98,Taretal99}.

The initial state is  an additional electron in the left dot with an energy
$\varepsilon_L$ well below the chemical potential of the left lead, the final state is
an additional electron in the right dot which then (keeping $T_c=0$) 
is lifted above the chemical potential of the right lead.
Such a transfer cycle of the open system (coupled to electron reservoirs) 
in the Coulomb blockade regime requires a Hilbert space ${\cal H}^{(3)}$ spanned by three states 
$|0\rangle$, $|L\rangle$, and $|R\rangle$ \cite{SN96,BK99}. Here, 
the two basis states
$|L\rangle=|N+1,M\rangle$ and $|R\rangle=|N,M+1\rangle$ describe one additional
electron in the left (right) dot above a ground state $|0\rangle=|N,M\rangle$
(`empty state').

For the remainder of this section, we only consider the first part of the transfer cycle, i.e., 
the dynamics of the double dot isolated from the leads, and turn to the full
transfer cycle including tunneling to and from the leads in section \ref{tunnelsection}.

\subsection{Coherent Adiabatic Transfer}
An adiabatic transfer from left to right requires the simultaneous change of at least two parameters such as
\begin{eqnarray}\label{etpulse}
\varepsilon(t)&=&\varepsilon_0+\varepsilon_1 \cos \Omega t \nonumber\\
T_c(t)&=&-T_c \exp [-(t - t_0)^2/\tau^2].
\end{eqnarray}
This corresponds to a change of $\varepsilon(t)$ with a 
simultaneous switching  of the tunnel coupling $T_c(t)$ between the dots. The precise form of the pulse,
Eq.(\ref{etpulse}),
is not important and has been chosen for convenience here.

The instantaneous, hybridized eigenstates of the isolated coupled quantum dot 
are
\begin{eqnarray}
  \label{eq:twobytworesult}
|\pm \rangle &=& \frac{1}{N_{\pm}}\left[
\pm 2T_c |L\rangle + (\Delta \mp \varepsilon) |R\rangle \right]\\
N_{\pm}&:=& \sqrt{4|T_c|^2+(\Delta \mp \varepsilon)^2},\quad \Delta:=\sqrt{\varepsilon^2+4|T_c|^2}\nonumber.
\end{eqnarray}
The eigenvalues 
$\varepsilon_{\pm}=\pm \frac{1}{2}\Delta$ 
of the coupled system  represent two energy surfaces over the $T_c$-$\varepsilon$ plane, the lower
of which (ground state) is shown in Fig. 1a. The pulse Eq. (\ref{etpulse})
corresponds to a curve on the $\varepsilon_-$ surface. 
The corresponding change of the `inversion' $\langle \sigma_z \rangle_t$ is obtained 
from a numerical integration
of the (coherent) equation of motion 
for the density matrix of the system.
The result is shown in Fig. 2, together with the form of the pulses Eq. (\ref{etpulse}). 

In accordance with the adiabatic theorem, the initial groundstate $|L\rangle$ of the system is 
rotated into the instantaneous superposition $|-\rangle$ of $|L\rangle$ and $|R\rangle$,
Eq. (\ref{eq:twobytworesult}), 
if the rotation is `slow', i.e., $\Omega,\tau^{-1},t_0^{-1}\ll \Delta/\hbar$.
In this case, the time enters as a parameter into  the state $|-\rangle$ 
which is used to calculate the approximate expectation value
\begin{eqnarray}
\label{adiabaticsz}
\langle \sigma_z \rangle_{\rm ad} = -\varepsilon(t)/\Delta(t),
\end{eqnarray}
which excellently reproduces the overall form of the numerically obtained 
$\langle \sigma_z \rangle_t$. The form Eq. (\ref{adiabaticsz}) corresponds to
Crisp's solution for the adiabatic following of an atom in a near resonance light pulse
\cite{Crisp73,Allen}.

The exact solution exhibits the expected
Rabi oscillations with frequency $\Delta(t)/\hbar$
around the adiabatic value, which are strongest when the tunnel coupling is fully switched on.
\begin{figure}[t]
\includegraphics[width=0.5\textwidth]{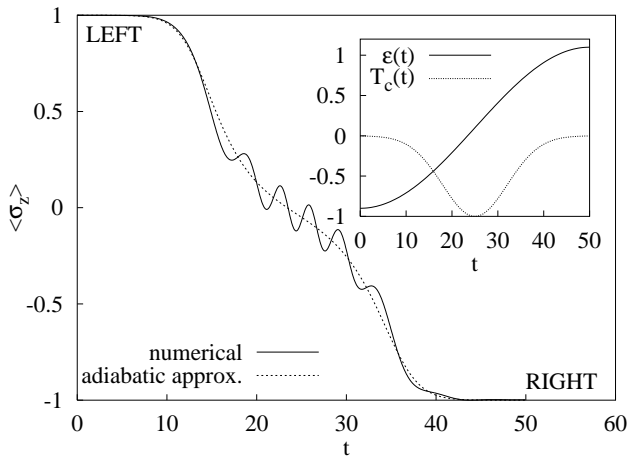}
\caption{\label{integro_2.eps}
Inversion $\langle \sigma_z \rangle$ for
transfer from left ($\langle \sigma_z \rangle=1$) to
right ($\langle \sigma_z \rangle=-1$)
in a two-level system, Eq. (\ref{H0define}). Inset: time-dependent 
tunnel matrix element $T_c(t)$ and energy splitting $\varepsilon(t)$, Eq. (\ref{etpulse}).
Energies (times) are in units of  the amplitude $T_c$ ($\hbar/T_c$) in Eq. (\ref{etpulse});
the other parameters are $t_0=25$, $\Omega= \pi/(2t_0)$, $\tau=10$, $\varepsilon_0=0.1$, $\varepsilon_1=-1$.
The adiabatic approximation Eq. (\ref{adiabaticsz})
is shown as dotted line.}
\end{figure}
Due to Landau-Zener tunneling from the adiabatic ground
state $|-\rangle$ to the excited state $|+\rangle$, there is always a finite albeit 
small probability $P_L$ for the electron to remain in the left dot, i.e., the excited state
after the rotation. $P_L$ can be made exponentially small for large enough level splitting $\Delta$
and slow pulses, but depends on the exact pulse shape $\varepsilon(t),T_c(t)$.

\subsection{Electron-Boson Coupling}
The adiabatic rotation discussed above is an idealization and valid only for a
two-level system completely isolated from its environment. In semiconductor quantum dots,
the coupling to low-energy bosonic excitations of the surrounding electron system, 
to photons, and to phonons leads to deviations from the coherent 
time-evolution. These excitations are characterized by
an effective spectral density
\begin{eqnarray}\label{rhodefinition}
\rho({\omega})&:=&\sum_{\bf Q} |g_{\bf Q}|^2\delta(\omega-\omega_{\bf Q}),
\end{eqnarray}
of modes ${\bf Q}$ with frequency $\omega_{\bf Q}$ and 
constants $g_{\bf Q}$ coupling to the electron charge in the dot. In this paper,
we only consider the coupling of phonons to the charge density 
which has been found to be dominant coupling mechanism in double quantum dots \cite{Fujetal98,Taretal99}. 
Other coupling mechanisms like inelastic spin-flip scattering \cite{HD00}
lead to additional dephasing channels, so that our results can only be considered as 
a lower bound on dephasing.
On the other hand, the time-dependent spin-boson Hamiltonian 
($\hbar=1$ throughout)
\begin{eqnarray}\label{modelhamiltonian}
  H^{(1)}(t) &=& H_0^{(1)}(t)+
\frac{1}{2}\sigma_z \hat{A} + H_B\\
\hat{A}&:=& \sum_{\bf Q} g_{\bf Q} \left(a_{-\bf Q} + a^{\dagger}_{\bf Q}\right), 
H_B:=\sum_{\bf Q}\omega_{Q} a^{\dagger}_{\bf Q} a_{\bf Q}\nonumber
\end{eqnarray}
employed here (and derived in Appendix A) is of sufficient general form
for our results being transferable to similar dissipative one-qubit rotations
in other two-level systems.

The function $\rho({\omega})$ can be calculated explicitely if one assumes sharply
peaked electron densities of negligeable width  
around the dot centers ${\bf r}_{L/R}$, having a distance
$d$ and zero extension into the $z$ (growth) direction. 
For bulk piezo-electric phonons, one obtains \cite{BK99}
\begin{eqnarray}\label{rhoomegadefinition}
\rho_{pz}({\omega}):= g \omega \left [1-\frac{\omega_d}{\omega}
\sin \left (\frac{\omega}{\omega_d} \right) \right] 
\theta(\omega),
\end{eqnarray}
where $g$ is a dimensionless coupling constant and $\omega_d=c/d$, with $c$ the velocity
of the phonon mode. 
In general, a finite extension $l$ of the electron densities in lateral or growth direction
leads to a form factor that cuts off phonons with frequencies $\omega \gtrsim l/c$
(`phonon bottleneck'). 

A microscopic determination 
of $\rho(\omega)$ would require (apart from details of the microscopic electron-phonon
interaction potential) exact knowledge of the many-body electron density in the dots.
We argue that the assumption of relatively sharply localized 
positions between which the additional electron tunnels is justified by the 
strong intra-dot electron-electron repulsion \cite{BV01}.
The origin of the oscillatory form of Eq. (\ref{rhoomegadefinition}) \cite{BK99}
lies in the double-slit like scattering of the phonons (waves with wave vector ${\bf Q}$)
at the electron charge that is delocalized between the two dots.

In the spin-boson problem, 
one often assumes a parametrized form of the spectral density \cite{g_and_alpha},
\begin{eqnarray}\label{rhosdefinition}
\rho_s({\omega}) := g \omega^s e^{-\frac{\omega}{\omega_c}} \theta(\omega),
\end{eqnarray}
where $\omega_c$ is a high-frequency cut-off. Notice that the ohmic case $s=1$
describes the microscopic form $\rho_{pz}({\omega})$, Eq. (\ref{rhoomegadefinition}),
in the limit $\omega_d/\omega\to 0$. 
However, phonons interacting with the dot through other coupling mechanisms 
like the deformation potential coupling, surface acoustic waves \cite{VBK01}, or
Rayleigh-Lamb waves in confined geometries \cite{DBK02}, lead to other forms of the spectral
density $\rho(\omega)$. Throughout the discussion of the one-qubit dynamics, 
we will assume the simple ohmic form, Eq.(\ref{rhosdefinition}) with $s=1$, 
and return to the form Eq. (\ref{rhoomegadefinition}) in the discussion of the 
two-qubit in section \ref{section_two_qubit}.

\subsection{Equations of Motion in Presence of Dissipation}
The coupling to the bosonic bath introduces decoherence and in general (exceptions
are discussed below) leads to a loss of fidelity of the adiabatic rotation, i.e. even for very slow
and adiabatic pulses $\varepsilon(t),T_c(t)$, the transfer from the left to the right state 
remains imperfect with the final expectation value $\langle \sigma_z \rangle$ considerably deviating
from $-1$. During the rotation, the pure adiabatic state decays into a mixture.

We solve for the reduced density matrix $\rho(t)$ 
of the two-level system  coupled to
external electron reservoirs \cite{SN96,Gur98,BK99}.
Different techniques can be applied for 
weak boson coupling (perturbation theory) and 
strong boson coupling (polaron transformations or
path integral in `NIBA' approximation \cite{GH98,HGGH00}).
In general, to obtain the solution for   
ime-dependent spin-boson problems even numerically is a non-trivial task \cite{GH98}.

From the Liouville-von Neumann equation, the diagonal elements of $\rho(t)$ are easily obtained
as
\begin{eqnarray}\label{rhoLLRR}
  \frac{\partial}{\partial t}\rho_{LL}(t)&=&-iT_c(t)\left[
\rho_{LR}(t)-\rho_{RL}(t) \right] \\
&+&
\gamma_L\left[1-\rho_{LL}(t)-\rho_{RR}(t)\right]
-\bar{\gamma}_L\rho_{LL}(t)\nonumber\\
\frac{\partial}{\partial t}\rho_{RR}(t)&=&-iT_c(t)\left[
\rho_{RL}(t)-\rho_{LR}(t) \right] \\
&+&
\gamma_R\left[1-\rho_{LL}(t) - \rho_{RR}(t)\right]
-\bar{\gamma}_R\rho_{RR}(t)\nonumber,
\end{eqnarray}
where we have included the coupling to the electron reservoirs (see Fig. 1) with
the tunnel rates ($j=L,R$)
\begin{eqnarray}
  \gamma_j&:=& \Gamma_j f_j(\varepsilon_j), \quad 
  \bar{\gamma}_j:= \Gamma_j\left[1- f_j(\varepsilon_j)\right],
\end{eqnarray}
where $f_j$ is the Fermi distribution in lead $j$ and $\Gamma_j$ the rate for tunneling between
dot $j$ and lead $j$. 

The equations for the off-diagonal elements $\rho_{RL}(t)=\rho_{LR}^*(t)=\langle L | \rho(t) |R \rangle$ 
can be obtained only approximatively by performing a perturbation expansion in
either $g$ or $T_c$:

\subsubsection{Born-Markov Approximation}
In the first approach, one starts from the basis of the hybridized states $|\pm \rangle$ 
and considers the term $\hat{V}:=\frac{1}{2}\hat{A}\sigma_z$ in the Hamiltonian Eq.(\ref{modelhamiltonian})
as a weak perturbation. If $\varepsilon$ and $T_c$ were constant, one could easily define
an interaction picture with respect to $H_0:= H-\hat{V}$ and proceed in the standard
way, i.e. second order perturbation theory and tracing out of the bosonic degrees of freedoms.
However, when $\varepsilon(t)$ and $T_c(t)$ become a function of time,
the time-evolution of the unperturbed system in general can not be written down analytically which makes the 
evaluation of the electron-boson terms very tedious. 

Here, we use an adiabatic approximation by regarding the time $t$ in $\varepsilon(t)$ and $T_c(t)$
as a parameter for the derivation of the incoherent (electron-boson) part of the master equation for 
$\rho_{LR}(t)$.
We neglect memory effects of the bosonic system and derive $\rho_{LR}(t)$ up to second order in
$\hat{V}$ (Born-Markov approximation).
The bosonic environment enters solely via the correlation function of the operator $\hat{A}$, 
Eq. (\ref{modelhamiltonian}),
in the interaction picture,
$K(t)= \langle \tilde{A}(t) \tilde{A}(0) \rangle$ which yields
\begin{equation}
K(t) 
= \int_0^{\infty}\!d\omega \, \rho(\omega)
     [n_B(\omega) e^{i\omega t} + (1\!+\!n_B(\omega)) \, e^{-i \omega t}],
\end{equation}
where $n_B(\omega)=[e^{\beta \omega}-1]^{-1}$ is the Bose distribution at temperature $T$
($\beta=1/k_B T$).
The result is
\begin{eqnarray}
\label{rhoLR}
\frac{d}{dt} \rho_{LR}(t) &=& \left( i \varepsilon(t) - 
\frac{\bar{\gamma}_L + \bar{\gamma}_R}{2} \right) \rho_{LR}(t) \\
&+& i T_c(t) \, (\rho_{RR}(t)-\rho_{LL}(t)) \nonumber\\
&-& \gamma(t) \,\rho_{LR}(t) + \gamma_+(t) \,\rho_{LL}(t) - \,\gamma_-(t) \rho_{RR}(t)\nonumber.
\end{eqnarray}
Here, the coefficients $\gamma$ and $\gamma_{\pm}$ (suppressing the 
parameter $t$ in the following) are defined as
\begin{eqnarray}\label{gammarates}
\gamma   &:=& \frac{1}{\Delta^2} \int_0^{\infty}\!\!dt \; (\varepsilon^2 
+ 4T_c^2 \cos \Delta t) \, \mbox{Re}\{K(t)\} ,\\
\gamma_+ &:=& \frac{T_c}{\Delta^2} \int_0^{\infty}\!\!dt \;
(\varepsilon \,(1\!-\!\cos \Delta t ) - i \Delta \sin \Delta t)\; K(t) ,\nonumber\\
\gamma_- &:=& \frac{T_c}{\Delta^2} \int_0^{\infty}\!\!dt \;
(\varepsilon \,(1\!-\!\cos\Delta t) - i \Delta \sin\Delta t)\; K^*(t) \nonumber.
\end{eqnarray}
The coupling to the bosonic bath 
introduces a dephasing rate $\gamma$ of the off-diagonal element $\rho_{LR}$,
\begin{eqnarray}
  \gamma =  2\pi\frac{T_c^2}{\Delta^2}\rho(\Delta) \coth\left(\frac{\beta \Delta}{2}\right),
\end{eqnarray}
and additional terms ($\gamma_{\pm}$) in the coupling to the diagonals.
In the ohmic case $s=1$ of the spectral density $\rho(\Delta)$ an additional
contribution to $\gamma$ linear in the temperature $T$ appears which, however,
is unphysical in the sense that for the microscopic $\rho_{pz}({\omega\to 0})\propto \omega^3$ 
this term does not appear, cf.  Eq.(\ref{rhoomegadefinition}) and
Appendix B. 
Note that in order to be consistent within the Born-Markov approximation, 
the scattering rates in Eq. (\ref{gammarates}) have to fulfill \cite{Bloch57}
$\gamma_{(\pm)} \ll k_BT$.

One should also keep in
mind that within our adiabatic approximation, these rates are time-dependent through
the time-dependence of $\varepsilon(t)$, $T_c(t)$, and
$\Delta=\Delta(t)=\sqrt{\varepsilon^2(t)+4T_c^2(t)}$.
Furthermore, the imaginary parts of $\gamma_{\pm}$
can be understood as a renormalization of the tunneling rate $T_c$, as can
be seen from Eq. (\ref{rhoLR}). At finite temperatures, they have to be computed numerically.

\subsubsection{Polaron Transformation}
One can derive an equation for the off-diagonal $\rho_{LR}(t)$ in a second, 
alternative approach by performing a unitary polaron transformation of the original Hamiltonian.
This method basically is a perturbation theory in the tunnel-coupling $T_c$ and suitable for
strong coupling to the bosonic bath. 
The unitary transformation
\begin{eqnarray}\label{polarontrafo}
  H\to \bar{H}:= e^{S}He^{-S}, S=\frac{1}{2}\sigma_z\sum_{\bf Q} \frac{g_{\bf Q}}{\omega_Q}
\left( a^{\dagger}_{\bf Q} - a_{-\bf Q}\right)
\end{eqnarray}
is applied to the Hamiltonian, leading to 
\begin{eqnarray}\label{Hpolaron}
  \bar{H}= \frac{\varepsilon(t)}{2}\sigma_z +  T_c(t) \left[
X\sigma_+ + X^+ \sigma_- \right] + H_B,
\end{eqnarray}
where $\sigma_+ = |L\rangle \langle R|$, $\sigma_- = |R\rangle \langle L|$, and a constant 
c-number energy term in $\bar{H}$ has been dropped. The transformation Eq.~(\ref{polarontrafo})
has removed the linear term $\frac{1}{2}\sigma_z \hat{A}$ and instead introduced the product
\begin{eqnarray}
  X =\Pi_{\bf Q}D_{\bf Q}\left(\frac{g_{\bf Q}}{\omega_{Q}}\right),\quad D_{\bf Q}(z):= e^{z a^{\dagger}_{\bf Q}
- z^* a_{\bf Q}}
\end{eqnarray}
of unitary displacement operators for the bosonic modes ${\bf Q}$ into the tunneling term. The 
master equation for $\rho_{LR}$ is now derived in this transformed picture in lowest order perturbation
theory in $T_c$. 
The result is
\begin{eqnarray}\label{eom3new}
\lefteqn{\rho_{LR}(t)= -\int_0^tdt'   e^{i\int_{t'}^{t}ds\,\varepsilon(s)}
\Bigg [ \frac{\bar{\gamma}_L+\bar{\gamma}_R}{2}C(t-t')
\rho_{LR}(t')} \nonumber\\
&\!+\!&
iT_c(t')  \left\{ C(t-t') \rho_{LL}(t') - C^*(t-t')
\rho_{RR}(t')\right\} \Bigg]
\end{eqnarray}
where $ C(t)$
is an equilibrium  correlation function with respect to the bosonic bath (inverse temperature
$\beta=1/k_BT$, spectral density $\rho(\omega)$, Eq. (\ref{rhodefinition})),
\begin{eqnarray}
 C(t)&:=&\langle X(t) X^{\dagger} \rangle_B  = e^{-\Phi(t)}\\
\Phi(t)&=& \int_0^{\infty}d\omega \frac{\rho(\omega)}{\omega^2} 
\left[ \left(1- \cos \omega t\right) \coth \left(\frac{\beta \omega}{2}\right)
+ i \sin \omega t \right].\nonumber
\end{eqnarray}

\subsection{Dissipative Adiabatic Transfer}
In order to discuss the effect of dissipation on the adiabatic transfer, we now numerically evaluate and 
compare the equations of motion for the density matrix $\rho(t)$ in both the perturbative and the polaron 
transformation 
approach. Here, we consider a closed two-level system without coupling to external electron reservoirs
($\Gamma_L=\Gamma_R=0$) and comment on the `open' system case (`quantum pump') below. 
The boson spectral density $\rho(\omega) = \rho_{s=1}(\omega)$, Eq. (\ref{rhosdefinition}),
always has ohmic form.

\begin{figure}[t]
\includegraphics[width=0.5\textwidth]{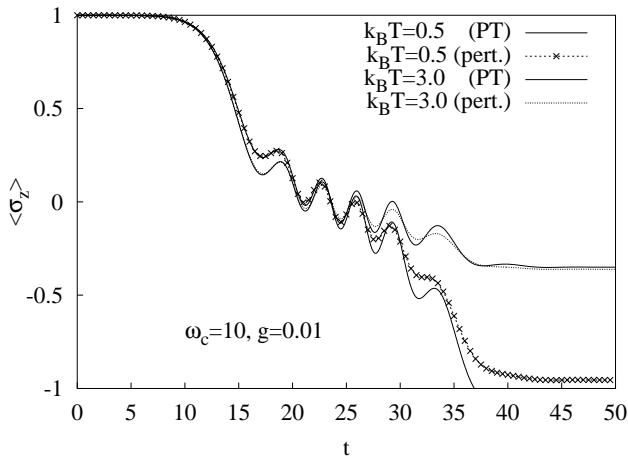}
\caption{\label{integro_3.eps}Two-level system as in Fig. (\ref{integro_2.eps}),
but with (weak) dissipation at finite temperatures $T$.
Comparison between the strong coupling polaron transformation (PT) solution and 
the perturbative solution is made.}
\end{figure}

First, the strong coupling equations Eq. (\ref{rhoLLRR}) and Eq. (\ref{eom3new}) can be
condensed  into a closed equation for
$\langle \sigma_z \rangle_t$ as
\begin{eqnarray}\label{polaronsigmaz}
  \frac{\partial}{\partial t}\langle \sigma_z \rangle_t 
&=&-\int_0^t dt' \sum_{\pm}
 \left[1\pm\langle \sigma_z \rangle_{t'}\right]f_{\pm}(t,t')\\
f_{\pm}(t,t')&:=& \pm 2T_c(t)T_c(t') {\rm Re } 
\left\{e^{\pm i\int_{t'}^t ds \varepsilon(s)} C(t-t')\right\}.\nonumber
\end{eqnarray}
This single integro-differential equation \cite{GH98} can be solved by standard numerical techniques.
Results from both the perturbative approach, Eq.(\ref{rhoLR}) and Eq.(\ref{rhoLLRR}), and
the polaron transformation, Eq.(\ref{polaronsigmaz}) are shown in Fig.(\ref{integro_3.eps}) 
for the weak coupling regime ($g=0.01$),
where the same form of pulses Eq.(\ref{etpulse}) as in the coherent case, Fig.(\ref{integro_2.eps}), 
has been chosen. 

We first checked
in a separate calculation that the influence of the imaginary parts $\mbox{Im}\{\gamma_{\pm}\}$, 
Eq. (\ref{raten-ohmsch}), on the solution of the master equation Eq. (\ref{rhoLR}) is negligible.
The perturbative inclusion of dissipation at low temperatures leads to the expected, 
small change of the time-evolution of $\langle \sigma_z \rangle$, with $\langle \sigma_z \rangle$
saturating  slightly above the  value $\langle \sigma_z \rangle=-1$ of the coherent case.
This means that in the dissipative case, the one-qubit rotation, i.e. the transfer of the electron from
left to right,  is never complete even if the rotation is performed adiabatically.
We will derive an analytic expression for this loss of fidelty in presence of dissipation 
in section \ref{sectionfidelity}.

In comparison to the perturbative approach, the polaron transformation approach
breaks down at weak couplings $g$ and low temperatures where it predicts inversions 
below $-1$ towards the end of the rotation. 
This breakdown if well-known from the static case ($T_c$, $\varepsilon$ fixed). On the other hand, the 
results from the polaron approach agree fairly well  with the perturbative
results at higher temperatures, 
as can be clearly infered from Fig. (\ref{integro_3.eps}) which again is consistent with the 
spin-boson dynamics for fixed $T_c$  and  $\varepsilon$.

\section{Fidelity of One-Qubit Swap}
\label{sectionfidelity}
The previous discussion has shown that the adiabatic transfer remains incomplete 
in the presence of dissipation,
i.e. $\langle \sigma_z \rangle$ can considerably deviate from its non-dissipative value
$-1$ in the perfectly adiabatic case where
no Landau-Zener transitions occur. 
Our aim in this section is to analyze how the deviation
\begin{eqnarray}
\delta\langle \sigma_z\rangle_t := \langle \sigma_z\rangle_t +1
\end{eqnarray}
depends on temperature,
boson spectral density $\rho(\omega)$, and the duration of the swap operation. 
Intuitively, it is clear that for too long swap duration, inelastic transitions to 
the excited level will have sufficient time to destroy the coherent transfer. On the other hand,
if the swap operation proceeds too fast, Landau-Zener transitions become stronger and
again lead to  deviations $\delta\langle \sigma_z \rangle>0$. This means
that for any given inelastic rate $\Gamma_{\rm in}$ there should be an optimal swap duration such
that these two competing processes balance each other.

One problem in quantifying this argument is that $\Gamma_{\rm in}$ naturally is time-dependent itself.
Silvestrini and Stodolsky nevertheless have suggested \cite{SS01} to extract
decoherence rates in experiments by
performing an adiabatic sweep and determining the maximum sweep duration for which the inversion
is still successfull. However, their model of a Bloch equation (with $\Gamma_{\rm in}$
as a constant parameter) is too simple in order to make detailed quantitative predictions here.

In the following, we calculate
the deviation $ \delta \langle \sigma_z \rangle_t$ for both weak and strong electron-boson coupling
and derive analytic expressions that allow to quantify the arguments above.
The main physical idea in this discussion is the introduction of
curves (ellipses in the $\varepsilon$-$T_c$ plane) where the excitation  energies
\begin{eqnarray}\label{ellipse}
  \Delta(t) = \sqrt{\varepsilon(t)^2+4T_c(t)^2}=\Delta
\end{eqnarray}
remain constant as a function of time and consequently, dissipation is determined 
by a  constant boson spectral density $\rho(\Delta)$.

\subsection{Transfer Rate: Exact Solution for Rabi Rotation and Weak Dissipation}
In the weak coupling case, 
we follow Grifoni and H\"anggi \cite{GH98} and perform a unitary transformation of  the original Hamiltonian (\ref{modelhamiltonian})
to a Hamiltonian $\bar{H}:=UHU^{-1}$, 
\begin{eqnarray}
  \bar{H}=-\frac{\Delta}{2}\tilde{\sigma}_z 
-\left(\frac{\varepsilon}{2\Delta}\tilde{\sigma}_z + \frac{T_c}{\Delta}
\tilde{\sigma}_x \right)\hat{A} + H_B,
\end{eqnarray}
where the matrix $U$ contains the columns of the 
hybridized eigenstates $|-\rangle$ and $|+\rangle$, Eq. (\ref{eq:twobytworesult}), 
$\tilde{\sigma}_z=|-\rangle\langle -| - |+\rangle\langle +|$, 
$\tilde{\sigma}_x=|-\rangle\langle +| + |+\rangle\langle -|$, 
and the time-dependence
of $T_c(t)$ and $\varepsilon(t)$ is again  considered as parametric.
We introduce the amplitude $a_{+-}(t)$ for an inelastic
transition from the adiabatic ground state $|-\rangle$ 
at time $t=0$ to the excited state $|+\rangle$ at time $t>0$, i. e. a transition induced
by the coupling to the bosons. The corresponding probability 
$P_{+-}(t)=|a_{+-}(t)|^2=\frac{1}{2}\delta \langle \sigma_z \rangle_t $ describes the deviation 
$\delta \langle \sigma_z \rangle_t$ 
of the inversion due to the coupling to the bosons.
The amplitude $a_{+-}(t)$  is given by the 
non-vanishing matrix element of the time evolution operator, expanded to lowest order,
\begin{eqnarray}
  a_{+-}(t)=i\int_0^tdt' \left\langle + \left| \frac{T_c(t')}{\Delta(t')}
\tilde{\sigma}_x(t')\tilde{A}(t') \right| -\right\rangle,
\end{eqnarray}
where the interaction picture is with respect to $-\Delta/2\tilde{\sigma}_z + H_B$.
With $\langle + |\tilde{\sigma}_x(t')|-\rangle$ = $\exp(i\int_0^{t'}ds \Delta(s))$,
we find the corresponding probability for a transition from $-$ to $+$ due to the 
interaction with the bosons as \cite{GH98}
\begin{eqnarray}\label{P+-}
   P_{+-}(t) &=& \int_0^{\infty}d\omega \rho(\omega) \Big\{
n_B(\omega) f(\omega,t) \nonumber\\
&+& \left[1+n_B(\omega)\right] f(-\omega,t)\Big\} \nonumber\\
f(\omega,t) &:=&\left|\int_0^t dt' \frac{T_c(t')}{\Delta(t')}
e^{-i\int_{0}^{t'}ds \left[\Delta(s)-\omega\right]}\right|^2.
\end{eqnarray}
`Elliptic' pulses $(T_c(t),\varepsilon(t))$, Eq. (\ref{ellipse}),
are defined by curves on the $\varepsilon_{-}$-surface (Fig. \ref{2surface})
with constant energy difference $\Delta$ to the excited state $\varepsilon_{+}$.
For the particular sinusoidal form 
\begin{eqnarray}\label{harmonicpulses}
  T_c(t) &=&-\frac{\Delta}{2}\sin \Omega t,\quad \varepsilon(t) = -\Delta \cos \Omega t
\end{eqnarray}
of $T_c(t)$ and $\varepsilon(t)$,
the time-dependent Hamiltonian $ H_0^{(1)}(t)= -(\Delta/2)[\cos(\Omega t)\sigma_z
+ \sin(\Omega t)\sigma_x]$,
Eq.(\ref{H0define}), becomes exactly integrable. It corresponds to
a spin $\frac{1}{2}$ in a magnetic field that rotates within the 
$x$-$z$-plane around the $y$-axis
with frequency $\Omega$. The solution for the inversion $\langle \sigma_z \rangle_t$ 
is easily obtained by transforming the Schr\"odinger equation into a rotating frame
(Rabi solution). One obtains
\begin{eqnarray}\label{sigmazrabi}
  \langle \sigma_z \rangle_t^{\rm Rabi}&=&
\left[ \left(\frac{\Delta}{\omega_R}\right)^2 +
\left(\frac{\Omega}{\omega_R}\right)^2 \cos\omega_R t \right]\cos\Omega t\\
&+& \frac{\Omega}{\omega_R}\sin\omega_R t\sin\Omega t,\quad \omega_R:=\sqrt{\Omega^2+\Delta^2}\nonumber.
\end{eqnarray}
Here, the Rabi frequency $\omega_R$ corresponds to maximal detuning $\delta = \Omega-0$ since
there is no static `magnetic field' $\propto \sigma_y$ in $y$-direction.

The use of the harmonic pulse Eq. (\ref{harmonicpulses}) has the further advantage that 
the quantity $f(\omega,t)$, Eq. (\ref{P+-}), can be evaluated analytically.
The swap operation requires
a pulse acting half a period from the initial time $t=0$ to the final time
$t_f=\pi/\Omega$. Using $t=t_f$ in $f(\omega,t)$, we obtain
\begin{eqnarray}
  \label{eq:fomega}
  f\left(\omega,\frac{\pi}{\Omega}\right)= 
\frac{1}{\Omega^2}\left[ \frac{\cos \left( \frac{\pi}{2}x \right)}
{x^2-1}\right]^2,\quad x := (\Delta-\omega)/\Omega.
\end{eqnarray}
In the adiabatic limit $\Omega/\Delta\to 0$, we find an approximation to the 
integrals Eq. (\ref{P+-}). In that limit, one has
\begin{eqnarray}
  f\left(\omega,\frac{\pi}{\Omega}\right) 
\to \frac{c}{\Omega}\delta(\Delta-\omega),\quad c = \frac{\pi^3J_{3/2}(\pi)}
{4\sqrt{2}}
\end{eqnarray}
whence the dissipation induced change of the inversion becomes
$\delta \langle \sigma_z \rangle_{f}^{\rm diss} = 2P_{+-}(t_f)=
2\frac{c}{\Omega}\rho(\Delta)n_B(\Delta)$.
The sum of the coherent contribution $\delta \langle \sigma_z \rangle_t$ from the 
Rabi solution at $t=t_f$, Eq. (\ref{sigmazrabi}), and the perturbative dissipative contribution
$\delta \langle \sigma_z \rangle_{f}^{\rm diss}$ leads to
\begin{eqnarray}\label{sigmazanalytic}
  \delta \langle \sigma_z \rangle_{f}&\approx& 
1-\left[ \left(\frac{\Delta}{\omega_R}\right)^2 +
\left(\frac{\Omega}{\omega_R}\right)^2 \cos\left(\frac{\pi \omega_R}{\Omega}\right) \right]\\
&+& 2\frac{c}{\Omega}\frac{\rho(\Delta)}{\exp({{\Delta}/{k_BT}})-1},\quad{\Omega}\ll{\Delta}, \quad c= 2.4674\nonumber.
\end{eqnarray}
\begin{figure}[t]
\includegraphics[width=0.5\textwidth]{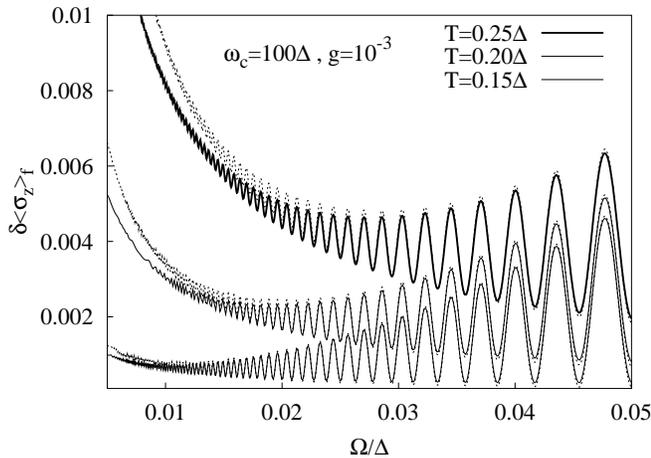}
\caption{\label{ddotzener_omega.eps}
Inversion change $\delta\langle \sigma_z\rangle_f$ after time $t_f=\pi/\Omega$ for 
sinusoidal pulses Eq.(\ref{harmonicpulses}) as 
obtained from the master equation, Eq.(\ref{rhoLR}).
Dotted curves correspond to the analytical prediction, Eq. (\ref{sigmazanalytic}).}
\end{figure}
Results for the inversion change $\delta \langle \sigma_z \rangle_{f}$ 
from the ideal value $\langle \sigma_z \rangle_{f}=-1$ as a function
of the pulse frequency $\Omega$ for a half-period sweep (duration $t_f=\pi/\Omega$) are shown
in Fig. (\ref{ddotzener_omega.eps}).
We compare the results from the (weak-coupling) master
equation,  Eq.(\ref{rhoLR}), and the analytical prediction Eq.(\ref{sigmazanalytic}).
The agreement between the numerical and the 
analytical prediction (dotted lines), is extremely good for small coupling constants ($g=10^{-3}$), 
but deviations become strong at small $\Omega$ where
the simple second order perturbation theory becomes worse.
The $1/\Omega$ dependence  of the dissipative contribution  to 
$\delta \langle \sigma_z \rangle_{f}$ is clearly visible at small $\Omega$, indicating that
for too long pulse duration the electron swap remains incomplete due to incoherent dissipation.
On the other hand, if the pulse duration is too short (larger $\Omega$), the oscillatory
coherent contribution from $\langle \sigma_z \rangle_f^{\rm Rabi}$ dominates.

One should bear in mind, however,
that Eq. (\ref{sigmazanalytic}) is an approximation that only holds in the limit of an
infinitely slow adiabatic change, i.e. $t_f=\pi/\Omega \to \infty$. 
In fact, 
for any finite pulse duration $t_f<\infty$, even in the 
limit of zero temperature $T=0$, Eq. (\ref{P+-}) yields
\begin{eqnarray}\label{P+-T=0}
  P_{+-}(t_f) = \int_0^{\infty}d\omega \rho(\omega) f(-\omega,t_f),
\end{eqnarray}
which shows that there is a small, but finite probability for inelastic processes during 
transitions from $|-\rangle$ to $|+\rangle$ even at zero temperature, in agreement
with \cite{GH98}.
These processes are due to the spontaneous emission of bosons which occur {\em during}
Landau-Zener transitions from $|-\rangle$ to $|+\rangle$ 
with a finite probability as long as $t_f$ is finite.

\subsection{Transfer Rate: Crossover for Strong Coupling}
For strong electron-boson coupling constants $g$, we used Eq. (\ref{polaronsigmaz})
to extract the final inversion $\langle \sigma_z \rangle_{f}$ after the time $t=t_f$ for 
the pulse Eq.(\ref{etpulse}). An interesting observation can be made in the temperature
behavior of $\langle \sigma_z \rangle_{f}$: for couplings $g\lesssim 2$, a temperature
increase leads to an increase of $\langle \sigma_z \rangle_{f}$, which is as in
the weak coupling case. However, above $g\gtrsim 2$, the temperature dependence changes in that
larger temperatures $T$ lead to smaller values of  $\langle \sigma_z \rangle_{f}$. 
In fact, for large coupling constants $g$, the system tends to remain localized 
in the left dot state $|L\rangle$ and no tunneling to the right state $|R\rangle$ occurs.
In this regime, higher temperatures destroy the localization and lead to smaller 
$\langle \sigma_z \rangle_{f}$. This behavior again is consistent with
the borderline $g=2$ ($\alpha=1$) \cite{g_and_alpha} 
in the dissipative two-level dynamics \cite{Legetal87} for static parameters $\varepsilon$ and $T_c$.

\begin{figure}[t]
\includegraphics[width=0.5\textwidth]{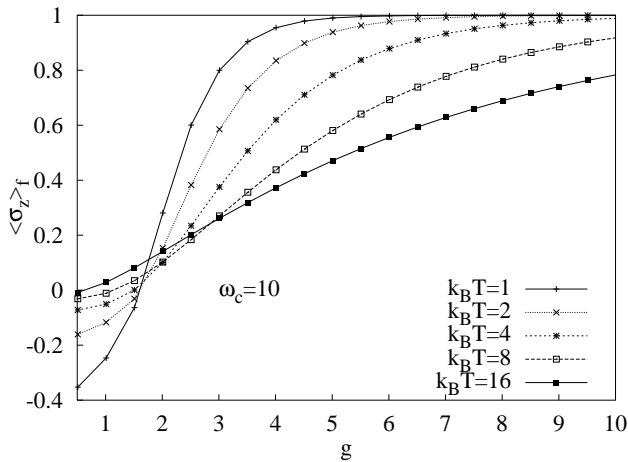}
\caption{\label{integro_4.eps}Inversion $\langle \sigma_z\rangle$ 
for strong electron-boson coupling after application of the 
pulse $T_c(t)$ and $\varepsilon(t)$, Eq. (\ref{etpulse}), 
as in Figs. (\ref{integro_2.eps}) and (\ref{integro_3.eps}).
Clearly visible is the crossover at $g\approx 2$ where the temperature dependence changes.}
\end{figure}

\subsection{Adiabatic Quantum Pumping in Open Double Dots}
\label{tunnelsection}
In the remainder of this section, we discuss how results for the 
closed double dot system, and in particular the 
coherent and incoherent inversion swap deviation $\delta \langle \sigma_z \rangle_{f}$ like the 
one for the Rabi rotation, 
Eq. (\ref{sigmazanalytic}), 
relate to electron transport in a double dot system coupled to 
external leads. Here, we only consider the weak dissipation case.

\begin{figure}[t]
\includegraphics[width=0.5\textwidth]{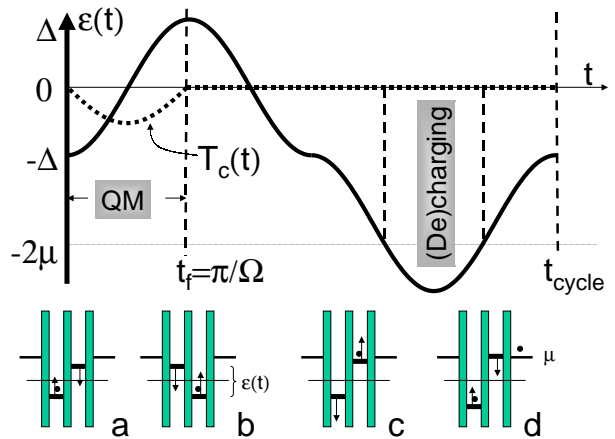}
\caption{\label{rabi.eps}Adiabatic scheme for cycling electrons from 
left to  right through a double quantum dot with
energy level difference $\varepsilon(t)=\varepsilon_L(t)-\varepsilon_R(t)$.
Left and right energy levels are assumed to change symmetrically 
with respect to $\varepsilon=0$.}
\end{figure}

The main idea is to apply time-dependent pulses such that
the quantum mechanical time evolution of 
the two-level system is well separated from a merely `classical' decharging and charging process.
One complete cycle of such an operation is sketched in Fig. \ref{rabi.eps}.
The cycle starts with an additional electron in the left dot and an adiabatic 
rotation of the parameters $(\varepsilon(t),T_c(t))$ such as, e.g.,  in Eq. (\ref{harmonicpulses})
or in Eq. (\ref{etpulse}), cf. Fig. \ref{rabi.eps} a-b. 
This completely quantum-mechanical part of the cycle is performed in
the `save haven' of the Coulomb- and the Pauli-blockade \cite{BR00}, i.e., with the left and right 
energy levels of the two dots well below the chemical potentials $\mu$ of the leads (which are
assumed identical here for simplicity). The cycle continues with closed 
tunnel barrier $T_c=0$ and increasing $\varepsilon_R(t)$ (Fig. \ref{rabi.eps} c); 
the two dots then are still in a superposition of the left and the right state.
The subsequent lifting of the right level above the chemical potential of
the right lead (Fig. \ref{rabi.eps} d) 
constitutes a measurement of that superposition (collapse of the wave-function):
the electron is
either in the right dot (with a high probability $1-\frac{1}{2}\delta\langle \sigma_z \rangle_f$)
and tunnels out, or the electron is in the left dot (and nothing happens because
the left level is still below $\mu$ and the system is Coulomb blocked).

If the tunnel rates $\Gamma_R,\Gamma_L$  to the right and left
leads are sufficiently larger than the inverse of the cyle duration $t_{\rm cycle}$, 
\begin{eqnarray}
  \Gamma_R,\Gamma_L \gg t_{\rm cycle}^{-1}
\end{eqnarray}
the decharging of the right dot and
the re-charging of the left dot from the left lead is fast enough to bring the system
back into its initial state with one additional electron on the left dot.
In this case, the precise value of  $\Gamma_R,\Gamma_L$, and the precise shape
of the $\varepsilon(t)$-pulse for $t_f<t<t_{\rm cycle}$  has no effect on
the total charge transfered within one cycle.
Then, since the probability to transfer one electron from the left to the right in one cycle is
given by $1-\frac{1}{2}\delta\langle \sigma_z \rangle_f$,
on the average an electron current 
\begin{eqnarray}\label{pumpcurrent}
  \langle I \rangle = -e \frac{1-\frac{1}{2}\delta\langle \sigma_z \rangle_f}{t_{\rm cycle}}
\end{eqnarray}
flows from left to right. Note that the leads 
essentially act as classical measurement devices
of the quantum-mechanical time-evolution between the two dots.
Measuring the current $\langle I \rangle$ as a function
of the pulse length $t_f=\pi/\Omega$ then offers a scheme to make quantum mechanical oscillations such as those
predicted in Eq. (\ref{sigmazanalytic}) visible in the electronic current,  similar to the recent experiment
by Nakamura {\em et al.} in a superconducting Cooper pair box \cite{NPT99}.

\subsection{Discussion}
The `Rabi-pulse', Eq. (\ref{harmonicpulses}),  keeps the energy difference to the excited state
$|+\rangle$ constant throughout the adiabatic rotation.
In this case, it follows that
the dissipative contribution to  $\delta \langle \sigma_z \rangle_{f}$,  Eq.(\ref{sigmazanalytic}),
is due to $\rho(\omega=\Delta/\hbar)$, i.e., one fixed boson frequency only. 
Furthermore, the analytic result, Eq.(\ref{sigmazanalytic}), in principle allows to
extract the value of the phonon spectral density, $\rho(\Delta)$, for `Rabi'-pulses
at fixed energy difference $\Delta$.

Another important observation is the fact that 
the rotation remains dissipation-free if 
$\Delta$ is chosen to coincide with a zero of $\rho(\omega)$. This defines a
`decoherence-free manifold' in the parameter space of the system.
Zeroes in $\rho(\omega)$ at certain frequencies $\omega=\omega_0$ have been predicted \cite{DBK02}
in free-standing, two-dimensional {\em phonon cavities} (slabs). 
In fact,  the control of vibrational
properties of quantum dot qubits has been suggested \cite{Fujetal98}, and 
considerable progress has been made in the fabrication of
nano-structures that are only partly suspended or even free-standing
\cite{CR98,BRWB98,Blietal00}.

In a thin-plate cavity model\cite{DBK02}, the zeroes in $\rho(\omega)$ are 
due to symmetry and geometrical confinement  both for
deformation potential and piezoacoustic phonon scattering in second order of the coupling 
constant $g$. 
This means that within a phonon cavity, the
dissipative contribution to $\delta \langle \sigma_z \rangle_{f}$, Eq. (\ref{sigmazanalytic}),
can be `switched off' if $\Delta$ is tuned to 
the energy $\hbar \omega_0$ of a decoupled phonon mode. 

For GaAs slabs of width $1 \mu$m,
this energy has been predicted to be of the order $\hbar\omega_0\approx 10 \mu$eV.
An energy $\Delta=10 \mu$eV corresponds to $k_BT \approx 100 $mK such that the
temperatures shown in Fig. \ref{ddotzener_omega.eps} are attainable in current experiments.
A typical pulse  frequency $\Omega/2\pi=0.01 \Delta /2\pi \hbar$ then would correspond to
$24$ MHz, the corresponding current being  of the order $\langle I \rangle \sim e\Omega/2\pi\approx
4 $pA. Furthermore, we mention that the weak coupling regime
with $g\approx 0.01\sim 0.05$ seems to be justified to describe recent experiments on phonon
coupling in double quantum dots \cite{Fujetal98,Taretal99}.

\section{Spin-Qubit Swap and Decoherence in Two-Qubits}
\label{section_two_qubit}
In this section, we apply our formalism to spin-qubit swaps and charge decoherence 
in two-qubit operations. These requires larger Hilbert spaces 
and can be realized by {\em two} electrons  on a double dot. 
Loss and DiVincenzo have introduced a detailed scheme for `quantum computation' with 
spin-states of coupled single-electron quantum dots \cite{LD98}. Dephasing of spin
degrees of freedom due to spin-orbit coupling or the coupling to nuclear spins is expected to be much weaker
than dephasing of charge degrees of freedom. Nevertheless, spin and charge become coupled 
during switching operations whereby charge dephasing also effects spin-based qubits \cite{LD98,BLD99}.

Here, we consider a specific two-qubit swap operation as
discussed recently by Schliemann, Loss and MacDonald: two 
electrons with spin  are localized on two coupled quantum dots $A$ and $B$, giving rise to a
basis of six states.  
During the operation {\em charge} decoherence occurs for intermediate states that 
get involved in the swap operation when charge is tunneling between the dots. 
Piezoelectric phonons then couple to the electron charge and 
incoherently mix states in the singlet sector which leads to a loss of fidelity of the swap operation. 
As in the previous section, we only consider the coupling of phonons to the charge density 
which yields a lower bound for dephasing.

\subsection{Two-Qubit Hamiltonian}
The four basis vectors with the two electrons on different dots are the spin singlet and triplets
\begin{eqnarray}
|S_1\rangle &:=&2^{-1/2}(c_{A\uparrow}^{\dagger}c_{B\downarrow}^{\dagger}
-c_{A\downarrow}^{\dagger}c_{B\uparrow}^{\dagger})|0\rangle\nonumber\\
|T^{-1}\rangle&:=&c_{A\downarrow}^{\dagger}c_{B\downarrow}^{\dagger}|0\rangle,\quad
|T^{1}\rangle:=c_{A\uparrow}^{\dagger}c_{B\uparrow}^{\dagger}|0\rangle\nonumber\\
|T^0\rangle &:=&2^{-1/2}(c_{A\uparrow}^{\dagger}c_{B\downarrow}^{\dagger}
+c_{A\downarrow}^{\dagger}c_{B\uparrow}^{\dagger})|0\rangle.
\end{eqnarray}
The remaining two states with two electrons on dot $A$ (`left') or dot $B$ (`right') are
\begin{eqnarray}\label{LRdefine}
  |L\rangle &:=& c^{\dagger}_{A\uparrow}c^{\dagger}_{A\downarrow}|0\rangle
= 2^{-1/2}\left[ |S_2\rangle + |S_3\rangle \right]\nonumber\\
  |R\rangle &:=& c^{\dagger}_{B\uparrow}c^{\dagger}_{B\downarrow}|0\rangle
= 2^{-1/2} \left[ |S_2\rangle - |S_3\rangle \right]
\end{eqnarray}
which are superpositions of two spin singlets $|S_{2,3}\rangle$,
\begin{eqnarray}
  |S_{2,3}\rangle &:=& 2^{-1/2} (c_{A\uparrow}^{\dagger}c_{A\downarrow}^{\dagger}
\pm c_{B\uparrow}^{\dagger}c_{B\downarrow}^{\dagger})|0\rangle
\end{eqnarray}
that differ in their orbital wave function.

As in \cite{SLM01}, we specify the  two-qubit swap as an adiabatic rotation 
from an initial state $|i\rangle$ to a final state $|f\rangle$,
\begin{eqnarray}\label{swapdefinition}
|i\rangle:=  \frac{1}{\sqrt{2}}\left[|T^0\rangle + |S_1\rangle\right]
\to |f\rangle :=\frac{1}{\sqrt{2}}\left[|T^0\rangle - |S_1\rangle\right]
\end{eqnarray}
that can be achieved \cite{LD98,BLD99} by an adiabatically opening and then closing of 
the tunnel barrier between the two dots as a function of time.
This operation leads out of the subspace span$\{|S_1\rangle,|T^0\rangle\}$ since it involves intermediate
doubly occupied states in  span$\{|L\rangle,|R\rangle\}$ (= span$\{|S_2\rangle,|S_3\rangle\}$) which
altogether define a four-dimensional Hilbert space, ${\cal H}^{(4)}$, in which the two-qubit swap
takes place. To be more specific \cite{BLD99}, we 
use a basis of ${\cal H}^{(4)}$ as given by the three singlets $|S_j\rangle$ and the triplet $|T^0\rangle$,
\begin{eqnarray}\label{spinbasis}
  |0\rangle:=|T^0\rangle,\quad |j\rangle:=|S_j\rangle,\quad (j=1,2,3),
\end{eqnarray}
to define the time-dependent Hamiltonian 
\begin{eqnarray}\label{H02define}
  H_0^{(2)}(t)&=&\sum_{j=0}^3\varepsilon_j |j\rangle \langle j|
+ T_c(t) \left[ |1\rangle \langle 2| +  |2\rangle \langle 1| \right],
\end{eqnarray}
where $\varepsilon_j$ denotes the energies of the spin singlet states, $\varepsilon_1=\varepsilon_0$, 
$\varepsilon_2=\varepsilon_0+U_H$, $\varepsilon_3=\varepsilon_0+U_H-2X$ with
the spin triplet energy $\varepsilon_0$, the on-site Coulomb repulsion $U_H>0$, 
the exchange term $X>0$, and the 
time-dependent tunnel coupling element between the dots $T_c(t)$.

\subsection{Electron-Boson Coupling, Master Equation}
The total Hamiltonian in presence of bosons coupling to the charge degree of freedom
is derived in Appendix B,
\begin{eqnarray}\label{modelhamiltonian2}
  H^{(2)}(t) &=& H_0^{(2)}(t)+
\frac{1}{2}\sigma_z \hat{A} + H_B,
\end{eqnarray}
it has exactly the same form as in the one-qubit case, Eq.(\ref{modelhamiltonian}),
but with the free Hamiltonian $H_0^{(1)}(t)$ replaced by $H_0^{(2)}(t)$, the 
coupling constants $g_{\bf Q}$ replaced by $\bar{g}_{\bf Q}$, 
and $\sigma_z:= |L\rangle\langle L|-|R\rangle\langle R|$ now referring to the 
two-particle states
Eq. (\ref{LRdefine}).
In analogy with the one-qubit case,
we introduce the matrix elements of the reduced density operator in the interaction picture
with respect to $H_p$ by
\begin{eqnarray}
  \rho_{ij}(t):={\rm Tr}_{\rm ph}\langle j | \tilde{\rho}(t)| i\rangle,\quad i = 0,1,L,R.  
\end{eqnarray}
The coherent part of the time evolution of the $\rho_{ij}(t)$ is obtained trivially
from the Liouville equation, using the Hamiltonian $H_0^{(2)}(t)$, Eq. (\ref{H02define}).
The non-trivial part are the
additional terms due to the electron-phonon coupling.
A systematic perturbation theory in the latter starts from the four adiabatic 
eigenstates of the unperturbed Hamiltonian $H_0^{(2)}(t)$, Eq. (\ref{H02define}). 

Here, we restrict ourselves to the case of small tunnel coupling,
\begin{eqnarray}\label{Tclimit}
 |T_c(t)| \ll U_H, 2X,
\end{eqnarray}
where inelastic transitions  are determined by the
dynamics in the subspace spanned by the states $|2\rangle$ and $|3\rangle$ and 
admixtures from $|1\rangle $  through the hybridization between $|1\rangle$ and
$|2\rangle$ can be neglected. This case is particularily interesting
since the (adiabatic) energy difference $2X$ between $|2\rangle$ and $|3\rangle$ remains constant
throughout the operation. As we show below, if $2X$  coincides with 
a zero  $\hbar\omega_0$ of the boson spectral density, the operation is again dissipation-less
in second order of the electron-boson coupling.

Within the Born and Markov approximation one finds
\begin{eqnarray}
  \left.\dot{\rho}_{LR}\right|_{\rm ep} &=& - \Gamma \rho_{LR} - \Gamma_- \rho_{RR} + \Gamma_+ 
\rho_{LL} \nonumber\\
 \left.\dot{\rho}_{Li}\right|_{\rm ep} &=& - \frac{\Gamma}{4} \rho_{Li} - \frac{\Gamma_-}{2} \rho_{Ri},
\quad i = 0,1 \nonumber\\
 \left.\dot{\rho}_{Ri}\right|_{\rm ep} &=& - \frac{\Gamma}{4} \rho_{Ri} + \frac{\Gamma_+}{2} \rho_{Li},
\quad i = 0,1.
\end{eqnarray}
Here, the rates $\Gamma$ and $\Gamma_{\pm}$ are defined in analogy with Eq.~(\ref{gammarates}),
\begin{eqnarray}\label{ratesdefinition2}
    \Gamma&=& 2\pi \frac{X^2}{\Delta^2} \rho (\Delta) \coth \left(\frac{\beta \Delta}{2}\right)\\
\Gamma_{\pm}&=& \mp  \frac{X}{\Delta} \frac{\pi}{2}\rho (\Delta),\nonumber
\end{eqnarray}
with the 
energy difference $\Delta$ given by
$
  \Delta  = 2X. 
$
Notice that  due to the limit Eq. (\ref{Tclimit}), $\Delta$ does not depend on time.

\begin{figure}[t]
\includegraphics[width=0.5\textwidth]{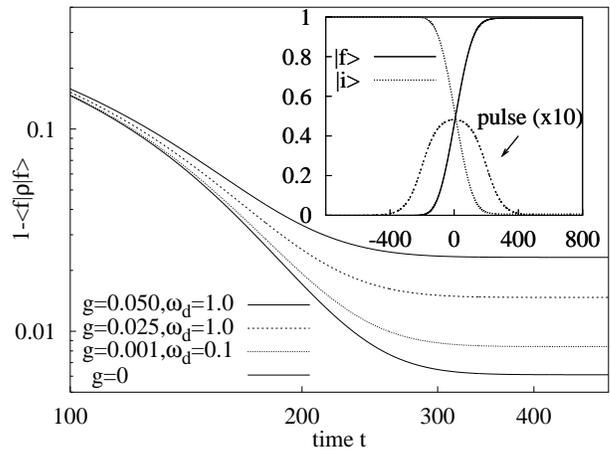}
\caption{\label{swap3.eps}Expectation value $1-\langle f |\rho(t)| f \rangle$
for a two-electron double dot qubit swap
from initial state $|i\rangle$ to final state $|f\rangle$, Eq.~(\ref{swapdefinition}),
as a function of time (in units of $U_H/\hbar$, $U_H$: onsite Coulomb repulsion). 
The corresponding dephasing times ($g>0$) are (from top to bottom)
$\Gamma^{-1}=   84$, $\Gamma^{-1}=  169$ and $\Gamma^{-1}= 635 $.
Inset: $\langle i |\rho(t)| i \rangle$ and $\langle f |\rho(t)| f \rangle$ for
vanishing phonon coupling $g=0$.}
\end{figure}

In Fig. (\ref{swap3.eps}), we show results of a numerical evaluation
of the two-electron master equation. The spectral density $\rho_{pz}(\omega)$, 
Eq. (\ref{rhoomegadefinition}), was used to obtain the rates, Eq.~(\ref{ratesdefinition2}),
for different coupling strengths $g$ and energies $\hbar\omega_d$
($\omega_d=c/d$, with $c$ the speed of sound and $d$ the dot center distance). 
Using $U_H=1$ as energy scale,
our results coincide with those of Schliemann et al.  for $g=0$ with a pulse of the form \cite{SLM01}
\begin{eqnarray}
  T_c(t) = \frac{T_0}{1+\cosh(t/\tau)/\cosh(T/2\tau)}
\end{eqnarray}
as shown in the inset. Here, we chose $T_0= 0.05$, $T=400$, 
$\tau = 50$, $X=0.5$, and the temperature $1/\beta=0.1$. 
A value of $U_H=1 $meV corresponds to a time of $4 \times 10^{-12}$ s.

The diagonal element $\langle f|\rho(t)|f\rangle$ is a measure of the fidelity of the 
swap operation, transfering $|i\rangle$ into $|f\rangle$, cf. Eq. (\ref{swapdefinition}).
Even in absence of dissipation, the non-adiabacity of the operation results in a finite value
of $1-\langle f|\rho(t)|f\rangle$ after the swap \cite{SLM01}. The electron-phonon
interaction acts when charge between the dots is moved during the opening
of the tunneling barrier. Consequently, the two states $|2\rangle$ and $|3\rangle$ become mixed
incoherently, leading to a finite, irreversible occupation probablity of the energetic lower
state $|3\rangle$ even after the pulse operation. 
We mention that the data shown here correspond to a worst case scenario where
the inelastic rate $\Gamma$, Eq. (\ref{ratesdefinition2}), is essentially determined by 
a large energy difference
$2X$ between $|2\rangle$ and $|3\rangle$ which enters into $\rho (\Delta=2X)$.
Then, spontaneous emission of phonons occuring during the slow swap (cf. Eq. (\ref{Tclimit}))
leads to a dephasing rate 
$\Gamma\approx \pi g X $. In this case, even relatively small values of $g$ ($g\lesssim 0.05$ in 
experiments on lateral dots \cite{Fujetal98,BK99}) can lead to a considerable fidelity loss
of the operation. 

On the other hand, our results also demonstrate that
a complete suppression (at least to lowest order in $g$), similar to the one-qubit case discussed
above, is theoretically possible for the two-qubit swap in phonon cavities. If $\Delta$ is tuned to 
the energy $\hbar \omega_0$ of a decoupled phonon mode, the inelastic rates, Eq. (\ref{ratesdefinition2}), 
can be `switched off' and the operation again is `dissipationless'.

\section{Conclusion}\label{conclusion}
In conclusion, we have studied  adiabatic quantum state rotations in coupled quantum dots
in presence of dissipation. For the one-qubit case, we suggested a scheme to extract
quantum oscillations in the inversion change $\delta \langle \sigma_z \rangle_{f}$
from the average pump current through the dots, Eq. (\ref{pumpcurrent}).
The analytic result, Eq.(\ref{sigmazanalytic}), furthermore  allows to
extract the value of the phonon spectral density, $\rho(\Delta)$, for `Rabi'-pulses
at fixed energy difference $\Delta$. 

For the two-qubit case, we have quantified the effect
of dissipation on a swap operation, where  states in the singlet sector are
incoherently mixed. Although we have only considered the case of small tunnel coupling between the dots,
as in the one-qubit case we have found that two-qubit rotations 
can be performed such that
dissipation is due to $\rho(\omega=\Delta/\hbar)$, i.e., one fixed boson frequency only. 
If $\Delta$ is chosen to coincide with a zero of $\rho(\omega)$
as in free-standing phonon cavities, this defines a
`decoherence-free manifold' in the parameter space of the system.

This work was supported by the EU via TMR and RTN projects
FMRX-CT98-0180 and HPRN-CT2000-0144, the German DFG project Br
1528/4-1, the UK project EPSRC GR44690/01 and
the UK Quantum Circuits Network. Discussions with R. H. Blick, S. De Franceschi, E. M. H\"ohberger,
L. P. Kouwenhoven, and F. Renzoni are gratefully acknowledged.

\begin{appendix}
\section{Charge-Phonon Coupling: One-Qubit}

Phonons coupling to the electron {\em charge} in a double quantum dot
are described by a spin-conserving electron-phonon Hamiltonian
\begin{eqnarray}
 H_{\rm ep} &=&  \sum_{\bf Q}\lambda_{\bf Q}\hat{\rho}_{\bf Q}\left[a_{-\bf Q} + a^{\dagger}_{\bf Q}\right],
\end{eqnarray}
where
$ a^{\dagger}_{\bf Q}$ is the creation operator for a phonon mode ${\bf Q}$, 
$\hat{\rho}_{\bf Q}$ the Fourier transform of the operator $\hat{\rho}({\bf x}) = \sum_\sigma
\Psi_{\sigma}^{\dagger}({\bf x})\Psi_{\sigma}({\bf x})^{\phantom{\dagger}}$ 
of the electron density, and 
$\lambda_{\bf{Q}} =\lambda_{-\bf{Q}}^*$ is the microscopic electron-phonon interaction matrix element.

The electron-phonon interaction potential in real space in first quantization is given by 
$  V_{\rm ep}({\bf{x}})=\sum_{\bf{Q}}\lambda_{\bf{Q}}e^{i\bf{Qx}}
\left(a_{-\bf{Q}}+a^{\dagger}_{\bf{Q}}\right)$.
We consider the simplest case of an electron-phonon interaction with only diagonal
terms 
\begin{eqnarray}\label{Hepab}
H_{\rm ep} &=& \sum_{\bf Q}\Big[ 
  \alpha_{\bf Q}^L\hat{N}_L 
+ \alpha_{\bf Q}^R\hat{N}_R \Big]\left[a_{-\bf Q} + a^{\dagger}_{\bf Q}\right],
\end{eqnarray}
where $\hat{N}_{i}=|i\rangle\langle i|$ ($i=L,R$),
$\alpha_{\bf Q}^{L/R}=\sum_\sigma\alpha_{\bf Q\sigma}^{L/R}$,
and 
\begin{eqnarray}\label{alphadef1}
\alpha_{{\bf Q}\sigma}^{L/R}&=&\lambda_{\bf Q} \int d^3{\bf x}e^{i{\bf Qx}}
\rho_{L/R,\sigma}({\bf x}).
\end{eqnarray}
Here, $\rho_{i,\sigma}({\bf x}):=\langle i | \Psi_{\sigma}^{\dagger}({\bf x})
\Psi_{\sigma}({\bf x})^{\phantom{\dagger}}|i\rangle$, $i=L,R$, 
is the density of electrons with spin $\sigma$ in the left (right) dot
as obtained from the (many-body) ground states $|L\rangle$ ($|R\rangle$).

Assuming two identical dots with 
$\rho_{i\sigma}({\bf x})=
\rho_{\sigma}({\bf x-r}_{L/R})$ to be smooth functions centered around the left (right) dot
centers ${\bf r}_{L/R}$, we obtain
\begin{eqnarray}\label{PQ}
\alpha_{\bf Q}^{L/R}&=&\lambda_{\bf Q} \exp\left({i{\bf Qr}_{L/R}}\right)P_e({\bf Q}),
\end{eqnarray}
where $P_e({\bf Q})=\sum_\sigma\int d^3{\bf x}e^{i{\bf Qx}}\rho_\sigma({\bf x})$ 
is the `form factor' of the electron density in the left and in the right dot.

In the electron-phonon interaction, Eq. (\ref{Hepab}), we write
\begin{eqnarray}\label{aplusb}
  \alpha_{\bf Q}^L \hat{N}_L + \alpha_{\bf Q}^R  \hat{N}_R =
\frac{1}{2}\left[ \sigma_z (\alpha_{\bf Q}^L- \alpha_{\bf Q}^R)
+  \hat{1}(\alpha_{\bf Q}^L + \alpha_{\bf Q}^R)\right]
\end{eqnarray}
and recognize that the term proportional to the  unity operator $\hat{1}$ in  the two-dimensional
Hilbert space ${\cal H}^{(2)}$ leads to a mere renormalization of the 
total energy by the phonons in Eq. (\ref{Hepab}): the energies $\varepsilon_{L/R}$ 
in both dots are shifted by the same amount. In fact,
if the coupling constants $\alpha_{\bf Q}^L$ and $\alpha_{\bf Q}^R$ were the same,
Eq.(\ref{Hepab}) would describe a trivial coupling of the phonons to 
$N_L+N_R=\hat{1}$  which would not affect the dynamics of the system, while here
(due to 
Eq. (\ref{PQ})) we find the phase relation 
\begin{equation}\label{phaseident2}
\alpha_{\bf{Q}}^R= \alpha_{\bf{Q}}^Le^{i{\bf{Qd}}},\quad
{\bf{d}}={\bf{r}}_R-{\bf{r}}_L,
\end{equation}
where $\bf{d}$ is the vector pointing from the center of the
left to the center of the right dot.

Defining $g_{\bf Q}=\alpha_{\bf Q}^L - \alpha_{\bf Q}^R$, the term
$\frac{1}{2}\sigma_z (\alpha_{\bf Q}^L- \alpha_{\bf Q}^R)$ in
Eq. (\ref{aplusb}) yields the interaction term $\frac{1}{2}\sigma_z \hat{A}$ in the
Hamiltonian $H^{(1)}(t)$, Eq. (\ref{modelhamiltonian}).

\section{The rates $\gamma$ and $\gamma_{\pm}$, Eq. (\ref{gammarates})}
The ohmic form Eq.(\ref{rhosdefinition}) with $s=1$ for the spectral function 
$\rho(\omega)=g \omega e^{-\omega/\omega_c}$ is used in the discussion of the one-qubit dynamics
in section \ref{section_one_qubit}. In this case, the corresponding rates  $\gamma$ and $\gamma_{\pm}$,
Eq. (\ref{gammarates}), are given by
\begin{equation}
\label{raten-ohmsch}
\begin{split}
\gamma &= \frac{g \pi}{\Delta^2} \left( \frac{\varepsilon^2}{\beta}
 + 2 T_c^2 \Delta e^{-\Delta/\omega_c} \coth\left(\frac{\beta \Delta}{2}\right) \right), \\
\mbox{Re}\{\gamma_{\pm}\} &= g \frac{\pi T_c}{\Delta^2}
\left( \frac{\varepsilon}{\beta} - \frac{\varepsilon}{2} \,\Delta\,
e^{-\Delta/\omega_c} \coth \left(\frac{\beta \Delta}{2}\right) \right.\\
&\mspace{200mu}\left.\mp \frac{\Delta^2}{2} \,e^{-\Delta/\omega_c} \right),\\
\mbox{Im}\{\gamma_{\pm}\} &= \mp g \, \frac{\varepsilon T_c \omega_c}{\Delta^2}
+ g \, \frac{T_c}{2 \Delta^2} \; \dashint_0^{\infty}\!\! d\omega \;\omega 
e^{-\omega/\omega_c} \\
& \mspace{50mu}\left\{ 
\frac{1}{e^{\beta \omega}-1} \left(\frac{\Delta\mp\varepsilon}{\omega-\Delta}
    - \frac{\Delta\pm\varepsilon}{\omega+\Delta}\right) \right. \\
& \mspace{60mu} \left. + \frac{1}{1-e^{-\beta\omega}}
\left( \frac{\Delta\pm\varepsilon}{\omega-\Delta}
    - \frac{\Delta\mp\varepsilon}{\omega+\Delta} \right) \right\}.
\end{split}
\end{equation}
Note that the ohmic case $s=1$ is peculiar in the sense that 
a contribution $\propto 1/\beta=T$ linear in the temperature $T$ appears in $\gamma$ and
$\mbox{Re}\{\gamma_{\pm}\}$ due to 
the limit $\lim_{\omega\to 0} \rho(\omega)n_B(\omega)$ under the integral in 
Eq. (\ref{gammarates}), which gives a finite contribution for $s=1$. 
However, the microscopic calculation \cite{BK99} 
of $\rho_{pz}(\omega)$, Eq.(\ref{rhoomegadefinition}), shows that 
the ohmic spectral density as an `envelope' of $\rho_{pz}(\omega)$ fails to be correct
for small $\omega$ where $\rho_{pz}({\omega\to 0})\propto \omega^3$ for piezoelectric phonons in double quantum dots.
The contribution to $\gamma$ and $\mbox{Re}\{\gamma_{\pm}\}$  linear in $T$ therefore is
unphysical but of minor relevance 
in the numerical calculations anyway, as we have checked.

\section{Charge-Phonon Coupling: Two-Qubit}
As in the one-qubit case, we start from an electron-phonon interaction Hamiltonian
with only diagonal terms 
\begin{eqnarray}\label{Hepab2}
H_{\rm ep} &=& \sum_{{\bf Q}\sigma;j=A,B}
  \alpha_{{\bf Q}\sigma}^j  c^{\dagger}_{j\sigma}c^{\phantom{\dagger}}_{j\sigma}
\left[a_{-\bf Q} + a^{\dagger}_{\bf Q}\right],
\end{eqnarray}
where the matrix element
\begin{eqnarray}\label{alphadef2}
  \alpha_{{\bf Q}\sigma}^{j}&:=&\lambda_{\bf Q} \int d^3{\bf x}e^{i{\bf Qx}}
\langle j | \Psi_{\sigma}^{\dagger}({\bf x})\Psi_{\sigma}({\bf x})^{\phantom{\dagger}}|j\rangle
\end{eqnarray}
is calculated with the single particle states $|j=A,B\rangle$.
We assume that the single electron densities in Eq.(\ref{alphadef2}) are
spin-independent, $ \alpha_{\bf Q\sigma}^{j}= \alpha_{\bf Q}^{j}$ whence
\begin{eqnarray}\label{Hepab3}
H_{\rm ep} &=& \sum_{\bf Q}\Big[ 
  \alpha_{\bf Q}^A  \hat{N}_{A}
+ \alpha_{\bf Q}^B  \hat{N}_{B}
 \Big]\left[a_{-\bf Q} + a^{\dagger}_{\bf Q}\right]
\end{eqnarray}
with the number operators $\hat{N}_{j}:=\sum_\sigma 
c^{\dagger}_{j\sigma}c^{\phantom{\dagger}}_{j\sigma}$.
The non-vanishing matrix elements of the number operators are
$ \langle j|\hat{N}_{A}| j \rangle = 1$,$\langle j|\hat{N}_{B}| j \rangle = 1$  where $j=0,1$,
$\langle L|\hat{N}_{A}| L \rangle = 2$, and
$\langle R|\hat{N}_{B}| R \rangle =2$.
Here, we used the states $|L/R\rangle := 2^{-1/2}\left[ |S_2\rangle \pm |S_3\rangle \right]$
with two electrons on the left (right) dot, Eq. (\ref{LRdefine}).
Using the completeness relation  in ${\cal H}^{(4)}$ yields
\begin{eqnarray}
& &(\alpha_{\bf Q}^A+\alpha_{\bf Q}^B)[|0\rangle\langle 0| + |1\rangle\langle 1|
+ 2 \alpha_{\bf Q}^A |L\rangle\langle L| + 2  \alpha_{\bf Q}^B |R\rangle\langle R|\nonumber\\
&=& \hat{1}(\alpha_{\bf Q}^A+\alpha_{\bf Q}^B) + (\alpha_{\bf Q}^A-\alpha_{\bf Q}^B)\sigma_z,
\end{eqnarray}
where again we defined $\sigma_z:= |L\rangle\langle L|-|R\rangle\langle R|$.

The total Hamiltonian now can be written as
\begin{eqnarray}\label{modelhamiltonian2a}
    H^{(2)}(t)&=&\sum_{j=0,1,L,R}\varepsilon_j |j\rangle \langle j|
+ \frac{T_c(t)}{\sqrt{2}}
\left[ |1\rangle \left( \langle L| + \langle R|\right) + H. c. \right]\nonumber\\
&+& X\sigma_x
+ \frac{1}{2}\hat{A}\sigma_z + H_B,
\end{eqnarray}
where 
$\hat{A}= \sum_{\bf Q} \bar{g}_{\bf Q} \left(a_{-\bf Q} + a^{\dagger}_{\bf Q}\right)$,
$\varepsilon_{L/R}= \frac{1}{2}\left(\varepsilon_2 + \varepsilon_3 \right)$ and
$\sigma_x= |L\rangle \langle R| + |R\rangle \langle L|$.
Comparing the coupling constant $g_{\bf Q}=\alpha_{\bf Q}^L- \alpha_{\bf Q}^R$ 
in the one-qubit case  and 
$\bar{g}_{\bf Q}=2(\alpha_{\bf Q}^A- \alpha_{\bf Q}^B)$ in
the two-qubit case,
and keeping in mind the definitions for $\alpha_{\bf Q}^{L/R}$ and $\alpha_{\bf Q}^{A/B}$,
Eq. (\ref{alphadef1}) and Eq. (\ref{alphadef2}), 
we realize that for spin-independent electron densities fulfilling
\begin{eqnarray}\label{denscondition}
\langle L | \Psi_{\sigma}^{\dagger}({\bf x})\Psi_{\sigma}({\bf x})^{\phantom{\dagger}}|L\rangle&=&
\langle A | \Psi_{\sigma}^{\dagger}({\bf x})\Psi_{\sigma}({\bf x})^{\phantom{\dagger}}|A\rangle\nonumber\\
\langle R | \Psi_{\sigma}^{\dagger}({\bf x})\Psi_{\sigma}({\bf x})^{\phantom{\dagger}}|R\rangle&=&
\langle B | \Psi_{\sigma}^{\dagger}({\bf x})\Psi_{\sigma}({\bf x})^{\phantom{\dagger}}|B\rangle  
\end{eqnarray}
we obtain identical coupling constants
$g_{\bf Q}=\bar{g}_{\bf Q}$: the spin sum in Eq. (\ref{alphadef1}) yields a factor $2$. 
An exception is the one-qubit with only one electron having a fixed spin (so that there is no spin sum).
In that case and with Eq. (\ref{denscondition}) holding, 
the one-qubit coupling $g_{\bf Q}$ is half the two-qubit coupling $\bar{g}_{\bf Q}$.
\end{appendix}



\begin{thebibliography}{54}
\expandafter\ifx\csname natexlab\endcsname\relax\def\natexlab#1{#1}\fi
\expandafter\ifx\csname bibnamefont\endcsname\relax
  \def\bibnamefont#1{#1}\fi
\expandafter\ifx\csname bibfnamefont\endcsname\relax
  \def\bibfnamefont#1{#1}\fi
\expandafter\ifx\csname citenamefont\endcsname\relax
  \def\citenamefont#1{#1}\fi
\expandafter\ifx\csname url\endcsname\relax
  \def\url#1{\texttt{#1}}\fi
\expandafter\ifx\csname urlprefix\endcsname\relax\def\urlprefix{URL }\fi
\providecommand{\bibinfo}[2]{#2}
\providecommand{\eprint}[2][]{\url{#2}}

\bibitem[{\citenamefont{{N.~C.~van der Vaart, S. F. Godjin, Y.~V.~Nazarov,
  C.~J.~P.~M.~Harmans, J.~E.~Mooij, L.~W.~Molenkamp, and
  C.~T.~Foxon}}(1995)}]{Vaartetal95}
\bibinfo{author}{\bibnamefont{{N.~C.~van der Vaart, S. F. Godjin,
  Y.~V.~Nazarov, C.~J.~P.~M.~Harmans, J.~E.~Mooij, L.~W.~Molenkamp, and
  C.~T.~Foxon}}}, \bibinfo{journal}{Phys. Rev. Lett.}
  \textbf{\bibinfo{volume}{74}}, \bibinfo{pages}{4702} (\bibinfo{year}{1995}).

\bibitem[{\citenamefont{{R.~H.~Blick, R.~J.~Haug, J. Weis, D. Pfannkuche, K. v.
  Klitzing, and K.~Eberl}}(1996)}]{Blietal96}
\bibinfo{author}{\bibnamefont{{R.~H.~Blick, R.~J.~Haug, J. Weis, D. Pfannkuche,
  K. v. Klitzing, and K.~Eberl}}}, \bibinfo{journal}{Phys. Rev. B}
  \textbf{\bibinfo{volume}{53}}, \bibinfo{pages}{7899} (\bibinfo{year}{1996}).

\bibitem[{\citenamefont{{T.~Fujisawa, T.~H.~Oosterkamp, W.~G.~van~der~Wiel,
  B.~W.~Broer, R.~Aguado, S.~Tarucha, and
  L.~P.~Kouwenhoven}}(1998)}]{Fujetal98}
\bibinfo{author}{\bibnamefont{{T.~Fujisawa, T.~H.~Oosterkamp,
  W.~G.~van~der~Wiel, B.~W.~Broer, R.~Aguado, S.~Tarucha, and
  L.~P.~Kouwenhoven}}}, \bibinfo{journal}{Science}
  \textbf{\bibinfo{volume}{282}}, \bibinfo{pages}{932} (\bibinfo{year}{1998}).

\bibitem[{\citenamefont{{R.~H.~Blick, D.~Pfannkuche, R.~J.~Haug,
  K.~v.~Klitzing, and K.~Eberl}}(1998)}]{Blietal98b}
\bibinfo{author}{\bibnamefont{{R.~H.~Blick, D.~Pfannkuche, R.~J.~Haug,
  K.~v.~Klitzing, and K.~Eberl}}}, \bibinfo{journal}{Phys. Rev. Lett.}
  \textbf{\bibinfo{volume}{80}}, \bibinfo{pages}{4032} (\bibinfo{year}{1998}).

\bibitem[{\citenamefont{{S.~Tarucha, T.~Fujisawa, K.~Ono, D.~G.~Austin,
  T.~H.~Oosterkamp, W.~G.~van der Wiel}}(1999)}]{Taretal99}
\bibinfo{author}{\bibnamefont{{S.~Tarucha, T.~Fujisawa, K.~Ono, D.~G.~Austin,
  T.~H.~Oosterkamp, W.~G.~van der Wiel}}}, \bibinfo{journal}{Microelectr.
  Engineer.} \textbf{\bibinfo{volume}{47}}, \bibinfo{pages}{101}
  (\bibinfo{year}{1999}).

\bibitem[{\citenamefont{{D.~Loss, D.~P.~DiVincenzo}}(1998)}]{LD98}
\bibinfo{author}{\bibnamefont{{D.~Loss, D.~P.~DiVincenzo}}},
  \bibinfo{journal}{Phys. Rev. A} \textbf{\bibinfo{volume}{57}},
  \bibinfo{pages}{120} (\bibinfo{year}{1998}).

\bibitem[{\citenamefont{{G.~Burkard, D.~Loss, and D. P.
  DiVincenzo}}(1999)}]{BLD99}
\bibinfo{author}{\bibnamefont{{G.~Burkard, D.~Loss, and D. P. DiVincenzo}}},
  \bibinfo{journal}{Phys. Rev. B} \textbf{\bibinfo{volume}{59}},
  \bibinfo{pages}{2070} (\bibinfo{year}{1999}).

\bibitem[{\citenamefont{{R. H. Blick and H. Lorenz}}(2000)}]{BL00}
\bibinfo{author}{\bibnamefont{{R. H. Blick and H. Lorenz}}},
  \bibinfo{journal}{Proceedings IEEE International Symposium on Circuits and
  Systems} \textbf{\bibinfo{volume}{II}}, \bibinfo{pages}{245}
  (\bibinfo{year}{2000}).

\bibitem[{\citenamefont{{T. H. Stoof and Yu. V. Nazarov}}(1996)}]{SN96}
\bibinfo{author}{\bibnamefont{{T. H. Stoof and Yu. V. Nazarov}}},
  \bibinfo{journal}{Phys. Rev. B} \textbf{\bibinfo{volume}{53}},
  \bibinfo{pages}{1050} (\bibinfo{year}{1996}).

\bibitem[{\citenamefont{Brandes and Kramer}(1999)}]{BK99}
\bibinfo{author}{\bibfnamefont{T.}~\bibnamefont{Brandes}} \bibnamefont{and}
  \bibinfo{author}{\bibfnamefont{B.}~\bibnamefont{Kramer}},
  \bibinfo{journal}{Phys. Rev. Lett.} \textbf{\bibinfo{volume}{83}},
  \bibinfo{pages}{3021} (\bibinfo{year}{1999}).

\bibitem[{\citenamefont{{X. Hu and S. Das Sarma }}(2000)}]{HD00}
\bibinfo{author}{\bibnamefont{{X. Hu and S. Das Sarma }}},
  \bibinfo{journal}{Phys. Rev. A} \textbf{\bibinfo{volume}{61}},
  \bibinfo{pages}{062301} (\bibinfo{year}{2000}).

\bibitem[{\citenamefont{{N.~H.~Bonadeo, J.~Erland, D.~Gammon, D.~Park,
  D.~S.~Katzer, and D.~G.~Steel}}(1998)}]{Bonetal98}
\bibinfo{author}{\bibnamefont{{N.~H.~Bonadeo, J.~Erland, D.~Gammon, D.~Park,
  D.~S.~Katzer, and D.~G.~Steel}}}, \bibinfo{journal}{Science}
  \textbf{\bibinfo{volume}{282}}, \bibinfo{pages}{1473} (\bibinfo{year}{1998}).

\bibitem[{\citenamefont{{J. Schliemann, D.~Loss, and A. H.
  MacDonald}}(2001)}]{SLM01}
\bibinfo{author}{\bibnamefont{{J. Schliemann, D.~Loss, and A. H. MacDonald}}},
  \bibinfo{journal}{Phys. Rev. B} \textbf{\bibinfo{volume}{63}},
  \bibinfo{pages}{085311} (\bibinfo{year}{2001}).

\bibitem[{\citenamefont{{T. Brandes, F. Renzoni, and R. H.
  Blick}}(2001)}]{BRB01}
\bibinfo{author}{\bibnamefont{{T. Brandes, F. Renzoni, and R. H. Blick}}},
  \bibinfo{journal}{Phys. Rev. B} \textbf{\bibinfo{volume}{64}},
  \bibinfo{pages}{035319} (\bibinfo{year}{2001}).

\bibitem[{\citenamefont{{M. Thorwart and P. H\"anggi}}(2002)}]{TH02}
\bibinfo{author}{\bibnamefont{{M. Thorwart and P. H\"anggi}}},
  \bibinfo{journal}{Phys. Rev. A} \textbf{\bibinfo{volume}{65}},
  \bibinfo{pages}{012309} (\bibinfo{year}{2002}).

\bibitem[{\citenamefont{{Y. Nakamura, Yu. A. Pashkin, and J. S.
  Tsai}}(1999)}]{NPT99}
\bibinfo{author}{\bibnamefont{{Y. Nakamura, Yu. A. Pashkin, and J. S. Tsai}}},
  \bibinfo{journal}{Nature} \textbf{\bibinfo{volume}{398}},
  \bibinfo{pages}{786} (\bibinfo{year}{1999}).

\bibitem[{\citenamefont{{H. Engel and D. Loss}}(2001)}]{EL01}
\bibinfo{author}{\bibnamefont{{H. Engel and D. Loss}}}, \bibinfo{journal}{Phys.
  Rev. Lett.} \textbf{\bibinfo{volume}{86}}, \bibinfo{pages}{4648}
  (\bibinfo{year}{2001}).

\bibitem[{\citenamefont{{L. J. Geerligs, V.F.~Anderegg, P.A.M.~Holweg,
  J.E.~Mooij, H.~Pothier, D.~Esteve, C.~Urbina, and
  M.H.~Devoret}}(1990)}]{Geretal90}
\bibinfo{author}{\bibnamefont{{L. J. Geerligs, V.F.~Anderegg, P.A.M.~Holweg,
  J.E.~Mooij, H.~Pothier, D.~Esteve, C.~Urbina, and M.H.~Devoret}}},
  \bibinfo{journal}{Phys. Rev. Lett.} \textbf{\bibinfo{volume}{64}},
  \bibinfo{pages}{2691} (\bibinfo{year}{1990}).

\bibitem[{\citenamefont{{L. P. Kouwenhoven, A. T. Johnson, N. C. van der Vaart,
  C. J. P. M. Harmans, and C. T. Foxon}}(1991)}]{Kouetal91}
\bibinfo{author}{\bibnamefont{{L. P. Kouwenhoven, A. T. Johnson, N. C. van der
  Vaart, C. J. P. M. Harmans, and C. T. Foxon}}}, \bibinfo{journal}{Phys. Rev.
  Lett.} \textbf{\bibinfo{volume}{67}}, \bibinfo{pages}{1626}
  (\bibinfo{year}{1991}).

\bibitem[{\citenamefont{{H.~Pothier, P.~Lafarge, C.~Urbina, D.~Esteve, and
  M.H.~Devoret}}(1992)}]{Potetal92}
\bibinfo{author}{\bibnamefont{{H.~Pothier, P.~Lafarge, C.~Urbina, D.~Esteve,
  and M.H.~Devoret}}}, \bibinfo{journal}{Europhys. Lett.}
  \textbf{\bibinfo{volume}{17}}, \bibinfo{pages}{249} (\bibinfo{year}{1992}).

\bibitem[{\citenamefont{{(Ed.) H. Grabert and M. H. Devoret}}(1991)}]{Grabert}
\bibinfo{author}{\bibnamefont{{(Ed.) H. Grabert and M. H. Devoret}}},
  \emph{\bibinfo{title}{Single Charge Tunneling}}, vol. \bibinfo{volume}{294}
  of \emph{\bibinfo{series}{NATO ASI Series B}} (\bibinfo{publisher}{Plenum
  Press}, \bibinfo{address}{New York}, \bibinfo{year}{1991}).

\bibitem[{\citenamefont{Stafford and Wingreen}(1996)}]{SW96}
\bibinfo{author}{\bibfnamefont{C.~A.} \bibnamefont{Stafford}} \bibnamefont{and}
  \bibinfo{author}{\bibfnamefont{N.~S.} \bibnamefont{Wingreen}},
  \bibinfo{journal}{Phys. Rev. Lett.} \textbf{\bibinfo{volume}{76}},
  \bibinfo{pages}{1916} (\bibinfo{year}{1996}).

\bibitem[{\citenamefont{{Ph.~Brune, C. Bruder, and H.
  Schoeller}}(1997)}]{BBS97}
\bibinfo{author}{\bibnamefont{{Ph.~Brune, C. Bruder, and H. Schoeller}}},
  \bibinfo{journal}{Phys. Rev. B} \textbf{\bibinfo{volume}{56}},
  \bibinfo{pages}{4730} (\bibinfo{year}{1997}).

\bibitem[{\citenamefont{{Q. Sung, J. Wang, and T. Lin}}(2000)}]{SWL00}
\bibinfo{author}{\bibnamefont{{Q. Sung, J. Wang, and T. Lin}}},
  \bibinfo{journal}{Phys. Rev. B} \textbf{\bibinfo{volume}{61}},
  \bibinfo{pages}{12643} (\bibinfo{year}{2000}).

\bibitem[{\citenamefont{{T. H. Oosterkamp, T. Fujisawa, W. G. van der Wiel, K.
  Ishibashi, R. V. Hijman, S. Tarucha, and L. P.
  Kouwenhoven}}(1998)}]{Oosetal98}
\bibinfo{author}{\bibnamefont{{T. H. Oosterkamp, T. Fujisawa, W. G. van der
  Wiel, K. Ishibashi, R. V. Hijman, S. Tarucha, and L. P. Kouwenhoven}}},
  \bibinfo{journal}{Nature} \textbf{\bibinfo{volume}{395}},
  \bibinfo{pages}{873} (\bibinfo{year}{1998}).

\bibitem[{\citenamefont{{R.~H.~Blick, D.~W.~van~der~Weide, R.~J.~Haug, and
  K.~Eberl}}(1998)}]{Blietal98a}
\bibinfo{author}{\bibnamefont{{R.~H.~Blick, D.~W.~van~der~Weide, R.~J.~Haug,
  and K.~Eberl}}}, \bibinfo{journal}{Phys. Rev. Lett.}
  \textbf{\bibinfo{volume}{81}}, \bibinfo{pages}{689} (\bibinfo{year}{1998}).

\bibitem[{\citenamefont{{A. W. Holleitner, H. Qin, F. Simmel, B. Irmer, R. H.
  Blick, J. P. Kotthaus, A. V. Ustinov, and K. Eberl}}(2000)}]{Holetal00}
\bibinfo{author}{\bibnamefont{{A. W. Holleitner, H. Qin, F. Simmel, B. Irmer,
  R. H. Blick, J. P. Kotthaus, A. V. Ustinov, and K. Eberl}}},
  \bibinfo{journal}{New Journal of Physics} \textbf{\bibinfo{volume}{2}},
  \bibinfo{pages}{2.1} (\bibinfo{year}{2000}).

\bibitem[{\citenamefont{{M. Switkes, C. M. Marcus, K. Campman, and A. C.
  Gossard}}(1999)}]{Swietal99}
\bibinfo{author}{\bibnamefont{{M. Switkes, C. M. Marcus, K. Campman, and A. C.
  Gossard}}}, \bibinfo{journal}{Science} \textbf{\bibinfo{volume}{283}},
  \bibinfo{pages}{1905} (\bibinfo{year}{1999}).

\bibitem[{\citenamefont{{P. W. Brouwer}}(1998)}]{Bro98}
\bibinfo{author}{\bibnamefont{{P. W. Brouwer}}}, \bibinfo{journal}{Phys. Rev.
  B} \textbf{\bibinfo{volume}{58}}, \bibinfo{pages}{R10135}
  (\bibinfo{year}{1998}).

\bibitem[{\citenamefont{{M. L. Polianski and P. W. Brouwer}}(2001)}]{PB01}
\bibinfo{author}{\bibnamefont{{M. L. Polianski and P. W. Brouwer}}},
  \bibinfo{journal}{Phys. Rev. B} \textbf{\bibinfo{volume}{64}},
  \bibinfo{pages}{075304} (\bibinfo{year}{2001}).

\bibitem[{\citenamefont{{J. N. H. J. Cremers and P. W. Brouwer}}(2002)}]{CB02}
\bibinfo{author}{\bibnamefont{{J. N. H. J. Cremers and P. W. Brouwer}}},
  \bibinfo{journal}{Phys. Rev. B} \textbf{\bibinfo{volume}{65}},
  \bibinfo{pages}{115333} (\bibinfo{year}{2002}).

\bibitem[{\citenamefont{{M. Moskalets and M. B\"uttiker}}(2001)}]{MB01}
\bibinfo{author}{\bibnamefont{{M. Moskalets and M. B\"uttiker}}},
  \bibinfo{journal}{Phys. Rev. B} \textbf{\bibinfo{volume}{64}},
  \bibinfo{pages}{201305} (\bibinfo{year}{2001}).

\bibitem[{\citenamefont{{E. R. Mucciolo, C. Chamon, and C. M. Marcus,
  cond-mat/0112157}}(2001)}]{MCM02}
\bibinfo{author}{\bibnamefont{{E. R. Mucciolo, C. Chamon, and C. M. Marcus,
  cond-mat/0112157}}} (\bibinfo{year}{2001}).

\bibitem[{\citenamefont{{ F. Renzoni and T. Brandes}}(2001)}]{RB01}
\bibinfo{author}{\bibnamefont{{ F. Renzoni and T. Brandes}}},
  \bibinfo{journal}{Phys. Rev. B} \textbf{\bibinfo{volume}{64}},
  \bibinfo{pages}{245301} (\bibinfo{year}{2001}).

\bibitem[{\citenamefont{{Yu. Makhlin, G. Sch\"on, and A.
  Shnirman}}(2001)}]{MSS01}
\bibinfo{author}{\bibnamefont{{Yu. Makhlin, G. Sch\"on, and A. Shnirman}}},
  \bibinfo{journal}{Rev. Mod. Phys.} \textbf{\bibinfo{volume}{73}},
  \bibinfo{pages}{357} (\bibinfo{year}{2001}).

\bibitem[{\citenamefont{{D.~V.~Averin}}(1998)}]{Ave98}
\bibinfo{author}{\bibnamefont{{D.~V.~Averin}}}, \bibinfo{journal}{Solid State
  Comm.} \textbf{\bibinfo{volume}{105}}, \bibinfo{pages}{659}
  (\bibinfo{year}{1998}).

\bibitem[{\citenamefont{Averin}(1999)}]{Ave99}
\bibinfo{author}{\bibfnamefont{D.~V.} \bibnamefont{Averin}}, in
  \emph{\bibinfo{booktitle}{Quantum computing and quantum communications}}
  (\bibinfo{publisher}{Springer}, \bibinfo{address}{Berlin},
  \bibinfo{year}{1999}), vol. \bibinfo{volume}{1509} of
  \emph{\bibinfo{series}{Lecture Notes in Computer Science}}, p.
  \bibinfo{pages}{413}.

\bibitem[{\citenamefont{{M.~Grifoni and P.~H\"anggi}}(1998)}]{GH98}
\bibinfo{author}{\bibnamefont{{M.~Grifoni and P.~H\"anggi}}},
  \bibinfo{journal}{Phys. Rep.} \textbf{\bibinfo{volume}{304}},
  \bibinfo{pages}{229} (\bibinfo{year}{1998}).

\bibitem[{\citenamefont{{S. Debald, T. Brandes, and B. Kramer}}(2002)}]{DBK02}
\bibinfo{author}{\bibnamefont{{S. Debald, T. Brandes, and B. Kramer}}},
  \bibinfo{journal}{Phys. Rev. B} \textbf{\bibinfo{volume}{66}},
  \bibinfo{pages}{Rxxxxx} (\bibinfo{year}{2002}).

\bibitem[{\citenamefont{{A. M. Childs, E. Farhi, and J.
  Preskill}}(2001)}]{CFP01}
\bibinfo{author}{\bibnamefont{{A. M. Childs, E. Farhi, and J. Preskill}}},
  \bibinfo{journal}{Phys. Rev. A} \textbf{\bibinfo{volume}{65}},
  \bibinfo{pages}{012322} (\bibinfo{year}{2001}).

\bibitem[{\citenamefont{{M. Crisp}}(1973)}]{Crisp73}
\bibinfo{author}{\bibnamefont{{M. Crisp}}}, \bibinfo{journal}{Phys. Rev. A}
  \textbf{\bibinfo{volume}{8}}, \bibinfo{pages}{2128} (\bibinfo{year}{1973}).

\bibitem[{\citenamefont{Allen and Eberly}(1987)}]{Allen}
\bibinfo{author}{\bibfnamefont{L.}~\bibnamefont{Allen}} \bibnamefont{and}
  \bibinfo{author}{\bibfnamefont{J.~H.} \bibnamefont{Eberly}},
  \emph{\bibinfo{title}{Optical Resonance and Two-Level Atoms}}
  (\bibinfo{publisher}{Dover}, \bibinfo{address}{New York},
  \bibinfo{year}{1987}).

\bibitem[{\citenamefont{{T. Brandes, T. Vorrath}}(2001)}]{BV01}
\bibinfo{author}{\bibnamefont{{T. Brandes, T. Vorrath}}}, in
  \emph{\bibinfo{booktitle}{Recent Progress in Many Body Physics}}, edited by
  \bibinfo{editor}{\bibnamefont{{R. Bishop, T. Brandes, K. Gernoth, N. Walet,
  and Y. Xian}}} (\bibinfo{publisher}{World Scientific},
  \bibinfo{address}{Singapore}, \bibinfo{year}{2001}), Advances in Quantum Many
  Body Theory.

\bibitem[{g_a({Note that our dimensionless coupling constant $g$ is related to
  the coupling constant $\alpha$ in the spin--boson literature by
  $g=2\alpha$.})}]{g_and_alpha}
 (\bibinfo{year}{{Note that our dimensionless coupling constant $g$ is related
  to the coupling constant $\alpha$ in the spin--boson literature by
  $g=2\alpha$.}}).

\bibitem[{\citenamefont{{T. Vorrath, T. Brandes, and B. Kramer}}(2001)}]{VBK01}
\bibinfo{author}{\bibnamefont{{T. Vorrath, T. Brandes, and B. Kramer}}}, in
  \emph{\bibinfo{booktitle}{Recent Progress in Many Body Physics}}, edited by
  \bibinfo{editor}{\bibnamefont{{R. Bishop, T. Brandes, K. Gernoth, N. Walet,
  and Y. Xian}}} (\bibinfo{publisher}{World Scientific},
  \bibinfo{address}{Singapore}, \bibinfo{year}{2001}), Advances in Quantum Many
  Body Theory.

\bibitem[{\citenamefont{{S. A. Gurvitz}}(1998)}]{Gur98}
\bibinfo{author}{\bibnamefont{{S. A. Gurvitz}}}, \bibinfo{journal}{Phys. Rev.
  B} \textbf{\bibinfo{volume}{57}}, \bibinfo{pages}{6602}
  (\bibinfo{year}{1998}).

\bibitem[{\citenamefont{{L. Hartmann, I. Goychuk, M. Grifoni, and P.
  H\"anggi}}(2000)}]{HGGH00}
\bibinfo{author}{\bibnamefont{{L. Hartmann, I. Goychuk, M. Grifoni, and P.
  H\"anggi}}}, \bibinfo{journal}{Phys. Rev. E} \textbf{\bibinfo{volume}{61}},
  \bibinfo{pages}{R4687} (\bibinfo{year}{2000}).

\bibitem[{\citenamefont{Bloch}(1957)}]{Bloch57}
\bibinfo{author}{\bibfnamefont{F.}~\bibnamefont{Bloch}},
  \bibinfo{journal}{Phys. Rev.}
  \textbf{\bibinfo{volume}{105}}(\bibinfo{number}{4}), \bibinfo{pages}{1206}
  (\bibinfo{year}{1957}).

\bibitem[{\citenamefont{{P. Silvestrini and L. Stodolsky}}(2001)}]{SS01}
\bibinfo{author}{\bibnamefont{{P. Silvestrini and L. Stodolsky}}},
  \bibinfo{journal}{Phys. Lett. A} \textbf{\bibinfo{volume}{280}},
  \bibinfo{pages}{17} (\bibinfo{year}{2001}).

\bibitem[{\citenamefont{{A. J. Leggett, S. Chakravarty, A. T. Dorsey, M. P. A.
  Fisher, A. Garg, and W. Zwerger}}(1987)}]{Legetal87}
\bibinfo{author}{\bibnamefont{{A. J. Leggett, S. Chakravarty, A. T. Dorsey, M.
  P. A. Fisher, A. Garg, and W. Zwerger}}}, \bibinfo{journal}{Review of Modern
  Physics} \textbf{\bibinfo{volume}{59}}(\bibinfo{number}{1}),
  \bibinfo{pages}{1} (\bibinfo{year}{1987}).

\bibitem[{\citenamefont{{T. Brandes and F. Renzoni}}(2000)}]{BR00}
\bibinfo{author}{\bibnamefont{{T. Brandes and F. Renzoni}}},
  \bibinfo{journal}{Phys. Rev. Lett.} \textbf{\bibinfo{volume}{85}},
  \bibinfo{pages}{4148} (\bibinfo{year}{2000}).

\bibitem[{\citenamefont{{A.~N.~Cleland, M.~L.~Roukes}}(1998)}]{CR98}
\bibinfo{author}{\bibnamefont{{A.~N.~Cleland, M.~L.~Roukes}}},
  \bibinfo{journal}{Nature} \textbf{\bibinfo{volume}{392}},
  \bibinfo{pages}{160} (\bibinfo{year}{1998}).

\bibitem[{\citenamefont{{R.~H.~Blick, M.~L.~Roukes, W.~Wegscheider,
  M.~Bichler}}(1998)}]{BRWB98}
\bibinfo{author}{\bibnamefont{{R.~H.~Blick, M.~L.~Roukes, W.~Wegscheider,
  M.~Bichler}}}, \bibinfo{journal}{Physica B} \textbf{\bibinfo{volume}{249}},
  \bibinfo{pages}{784} (\bibinfo{year}{1998}).

\bibitem[{\citenamefont{{R. H. Blick, F. G. Monzon, W. Wegscheider, M. Bichler,
  F. Stern, and M. L. Roukes}}(2000)}]{Blietal00}
\bibinfo{author}{\bibnamefont{{R. H. Blick, F. G. Monzon, W. Wegscheider, M.
  Bichler, F. Stern, and M. L. Roukes}}}, \bibinfo{journal}{Phys. Rev. B}
  \textbf{\bibinfo{volume}{62}}(\bibinfo{number}{24}), \bibinfo{pages}{17103}
  (\bibinfo{year}{2000}).

\end{thebibliography}

\end{document}